\documentclass[aps,pre,preprint,showpacs,superscriptaddress]{revtex4-1}
\usepackage{graphicx,color}
\usepackage{amssymb,amsmath}
\usepackage{xcolor, ulem}
\begin{document}
\title{Particle dynamics and spatial $\mathrm{e^-e^+}$  density structures at QED cascading in circularly polarized standing waves}

\author{A.V. Bashinov}
\affiliation{Institute of Applied Physics, Russian Academy of Sciences, 603950 Nizhny Novgorod, Russia}
\author{P. Kumar}
\affiliation{Department of Physics, University of Lucknow, 226007 Lucknow, India}
\author{A.V. Kim}
\affiliation{Institute of Applied Physics, Russian Academy of Sciences, 603950 Nizhny Novgorod, Russia}

\date{\today}

\begin{abstract}
We present a comprehensive analysis of longitudinal particle drifting in a standing circularly polarized wave at extreme intensities when quantum radiation reaction (RR) effects should be accounted for. To get an insight into the physics of this phenomenon we made a comparative study considering the RR force in the Landau-Lifshitz or quantum-corrected form, including the case of photon emission stochasticity. It is shown that the cases of circular and linear polarization are qualitatively different.  Moreover, specific features of particle dynamics have a strong impact on spatial structures of the electron-positron ($\mathrm{e^-e^+}$) density created in vacuum through quantum electrodynamic (QED) cascades in counter-propagating laser pulses. 3D PIC modeling accounting for QED effects confirms realization of different pair plasma structures.

\end{abstract}

\maketitle

\section{Introduction}

Development of ELI \cite{ELI}, Appolon 10 \cite{Appolon10}, XCELS \cite{ XCELSa} and other projects aimed at obtaining extreme laser fields stimulates fundamental and applied study of the  interaction of superintense laser radiation with matter. One of the features of this interaction is a decisive role of photon emission by electrons (positrons) and the corresponding radiation reaction effect. The electron motion changes drastically due to the impact of photon emission \cite{Tamburini_NuclInst}. As a result, for example, there may occur counterintuitive effects in a linearly polarized field, such as anomalous radiative trapping in a standing wave \cite{ART_Gonoskov} and radiative trapping in a traveling wave \cite{RT_Pukhov}. Moreover,  not only changes in particles' momentum but also quantum (stochastic) nature of the photon emission play an important role \cite{Duclous_PPCF2011,Neitz_PRL,Yoffe_NJP2015,Bashinov_PRE2015}. Another important feature is also that emitted hard photons with energies above 1 MeV in an extreme laser field can create electron-positron pairs through multiphoton Breit-Wheeler processes \cite{Breit_PR1934,Narozhny_JETP1965}. Eventually, both the distribution function of particles and field distribution can be modified substantially due to avalanche-like electron-positron pair production (electromagnetic cascade) \cite{Bell_PRL2008} and back reaction of the produced plasma \cite{Nerush_PRL2011}.

For efficient emission of hard photons and their decay into $\mathrm{e^-e^+}$ pairs the critical factor is a transverse field which a particle experiences in its rest frame. A simple case with a strong transverse field that can be realized in laboratory is two counter-propagating laser pulses. Most theoretical studies of $\mathrm{e^-e^+}$ generation were recently devoted to this case, which is also very instructive for understanding the main physical processes involved in such QED plasma behavior. Already in the first paper by Bell and Kirk \cite{Bell_PRL2008} a prolific pair production at intensities of 10$^{24}$W~cm$^{-2}$ for a 1 $\mu$m laser with circular polarization was shown. This case of field polarization is advantageous to the linear polarization case in that the electric field in antinode plane is a steadily rotating vector, whereas in the case of  linear polarization the electric field is oscillating in time, and longitudinal particle motions in standing field configuration are different. In the linearly polarized case particle escaping from the high field region can be suppressed by the anomalous radiative trapping (ART) mode \cite{ART_Gonoskov}, whereas in a circularly polarized standing wave particles are drifting longitudinally, as they are initially sitting on the top of the hump of ponderomotive potential (antinode region). Different spatial pair plasma structures were observed in modeling counter-propagating circularly polarized laser pulses \cite{Bashmakov_POP2014,Grismayer_POP2016,Jirka_PRE2016,Kostyukov_POP2016}.   

The goal of the present work is to study the types of spatial $\mathrm{e^-e^+}$ density structures that can be realized in vacuum through QED cascades in counter-propagating laser pulses with circular polarization. The qualitative difference of these structures compared to the case of linear polarization makes this study interesting and fundamentally important for understanding QED plasma dynamics in laser fields. We will consider in detail particle drifting in an inhomogeneous field, especially longitudinal drifting as the most important process of particle escape for counter-propagating pulses when standing wave configuration is formed. To get an insight into the physics we will first present long-term density distributions, showing that with radiation reaction effects only the normal radiative trapping (NRT) regime \cite{ART_Gonoskov,Lehmann_PRE} is realized, unlike the case of linear polarization when particles can be trapped in the vicinity of antinode in the ART regime. In NRT and ART regimes particles are attracted due to RR effect to electric field node or antinode regions of standing wave, respectively. Since QED cascades are mainly generated in the high-field region, we will once again consider particular trajectories in a rotating electric field. We will revisit the earlier works where different types of motion have been analyzed. Among them is a stationary trajectory representing a circle, which has a long way of study. It was first considered in Refs. \cite{Steiger_PRA1972,Zeldovich_UFN1975} taking into account radiation losses. Based on particle motion, the authors of  Ref.\cite{Steiger_PRA1972} made an attempt to derive dispersion relations in plasma with inverse Faraday effect taken into account. Later the electron motion was investigated with allowance for the Lorentz-Abraham-Dirac (LAD) force \cite{Rohrlich}. A  nonlinear Thomson scattering cross-section was found for different limiting cases: the so-called radiation-dominated when RR force is comparable with the Lorentz force and quantum \cite{Bulanov_PPR2004}. The stationary trajectory not only allows obtaining exact expressions for ponderomotive force and dielectric permittivity but also determines stationary nonlinear plasma-field structures, accounting for the LAD force \cite{Bashinov_PP2013}. Dispersion relation characteristics of stationary trajectory were modified considering quantum corrections to the Landau-Lifshitz (LL) force \cite{Zhang_NJP2015,Kostyukov_POP2016}. It is also important that at intensities approaching $10^{24}$W$/$cm$^2$ electromagnetic cascades start to be generated along this trajectory \cite{Bell_PRL2008}. Cascade growth rates were estimated with different accuracy in Refs.\cite{Elkina_PRSTAB2011,Nerush_POP2011,Grismayer_POP2016,Fedotov_PRL2010}. Based on our revision we will show that the stochastic nature of photon emission additionally slows down the rate of drifting to the electric field node due to strong perturbation of particle motion and generates a new effect of particle diffusion of quantum nature. However, analysis of QED cascade development in an inhomogeneous field should include all particle channels of escaping from the high-field region. This was done for some particular cases in Refs. \cite{Fedotov_PRA2014} for the transverse drift and  in \cite{Kirk_PPCF2009,Duclous_PPCF2011} considering numerically longitudinal particle motion from the electric field antinode to the node. We present a comprehensive analysis of particle drifting at extreme intensities from which quantitative dependences of escaping rates as a functions of field amplitude are obtained. Comparison of the pair production growth rates and the main particle loss rate connected with the longitudinal drifting shows that three modes of QED cascades may be formed in a standing circularly polarized wave, giving rise to density distributions peaked at the antinode or node or in both regions. This conclusion is confirmed by PIC simulations.

\section{Particle motion: long-term distribution}
\label{Sec2}
We first consider long-term density distribution of electrons initially uniformly distributed in a plane standing circularly polarized wave, with  the radiation reaction effect taken into account. Of course, this is a direct consequence of single-electron motions, but it allows understanding the asymptotic behavior of a particle ensemble. Such a consideration allows introducing ART and NRT regimes in a standing linearly polarized wave \cite{ART_Gonoskov}. 

Without loss of generality, assume that electric $\bf E$ and magnetic $\bf B$ fields may be written in the  form
\begin{gather}
\label{E}
\mathbf{E}=\mathrm{Re}(a\cos(y)(\mathbf{z}-i\mathbf{x})e^{it}),\\
\mathbf{B}=\mathrm{Re}(a\sin(y)(\mathbf{z}-i\mathbf{x})e^{it}).\label{B}
\end{gather}
The fields are normalized to $m\omega_l c/e$, where $\omega_l$ is laser frequency, $m$ and $-e$ are the mass and charge of the electron, $c$ is the velocity of light, the $y$ axis is perpendicular to $\bf E,B$, $y$ and $t$ are normalized to $c/\omega_l$ and $1/\omega_l$, respectively. The equations of motion make an autonomous system:
\begin{gather}
\frac{dp_{\|}}{dt}=\frac{ap_{\bot}\sin(y)\sin(\varphi)}{\gamma}-F_{r}p_\|,\label{Plong}\\
\frac{dp_\bot}{dt}=-a\cos(y)\cos(\varphi)-\frac{ap_\|\sin(y)\sin(\varphi)}{\gamma}-F_{r}p_\bot,\label{Ptr}\\
\frac{d\varphi}{dt}=-1+\frac{a}{p_\bot}\left(\cos(y)\sin(\varphi)-\frac{p_\|}{\gamma}\sin(y)\cos(\varphi)\right),\label{Phi}\\
\frac{dy}{dt}=\frac{p_\|}{\gamma},\label{Yc}
\end{gather}
where dimensionless variables are used, $p_\|$ is the electron momentum along the $y$ axis, $p_\bot$ is the magnitude of momentum projection on the $xz$ plane, $\varphi$ is the angle between $\bf E$ and momentum projection on the $xz$ plane counterclockwise measured from $\bf E$, and $\gamma$ is electron Lorentz-factor. The momentum is normalized to $mc$. Momentum projections on the $x$ and $z$ axes are $p_x=p_\bot\sin(t+\varphi)$ and $p_z=p_\bot\cos(t+\varphi)$, respectively. $F_r$ is the factor of radiation reaction force $\bf F_{rr}$, so that ${\bf F_{rr}}=-{\bf p}F_r$. $F_r$ can be considered within the framework of different approaches.
\begin{enumerate}
	\item Without radiation reaction force
	\begin{equation}
	F_r=0.
	\label{Fr0}
	\end{equation}
	\item Radiation reaction force in the form of Landau-Lifshitz force (the main term proportional to $\gamma^2$  \cite{Tamburini_NJP}): 
	\begin{equation}
	\label{FLL}
	F_r=2\alpha\eta a^2\left[\cos^2(y)+p_\|^2+p_\bot^2\sin^2(\varphi)\right]/(3\gamma),
	\end{equation}
\end{enumerate}
	where $\alpha$ is fine structure constant, $\eta=\frac{\hbar\omega_l}{mc^2}$, and $\hbar$ is Planck constant. We omit here introduction of LAD force, since it was described many times in the previous works and gives the same results as the LL force, while both of them are valid in the range of  field frequency and field strength  parameters \cite{Bulanov_PRE2011}. 
	In the ultrarelativistic case, radiation power P is related to $F_r$ by  $P\approx F_rp^2/\gamma$ to an  accuracy of $1/\gamma^2$. Following \cite{BayerKatkov}, we introduce the radiation reaction force with quantum corrections.
\begin{enumerate}
\setcounter{enumi}{2}
	\item Radiation reaction force taking into account quantum corrections
	\begin{equation}
	\label{FrrQ}
	F_r=\frac{\alpha}{3\sqrt{3}\pi\eta\gamma}\int_0^\infty u\frac{4u^2+5u+4}{(1+u)^4}K_{2/3}(2u/3\chi)du,
	\end{equation}
\end{enumerate}
	where $K_\nu(x)$ is the modified Bessel function of the second kind of order $\nu$, and quantum parameter \cite{Nikishov_85,Ritus_85}
	\begin{equation}
	\label{Chi}
	\chi=a\eta\sqrt{\cos^2(y)+p_\|^2+p_\bot^2\sin^2(\varphi)}.
	\end{equation}
	The approximation of (\ref{FrrQ}) can be found in Sec.\ref{AppFr}.\par
	One more way of describing radiation losses is to use the quasiclassical approach \cite{BayerKatkov}. Particle motion between two acts of photon emission is described by equations without radiation reaction force $F_r=0$, and at the instant of emission the particle momentum decreases proportionally to the emitted photon momentum. This approach is modelled within the framework of the Monte-Carlo method \cite{Duclous_PPCF2011,Nerush_PRL2011,Gonoskov_PRE2015}. In our article we use the method described in Ref.\cite{Bashinov_PRE2015}.\par
\begin{figure}[h]
	\centering
		\includegraphics[width = 0.23\textwidth]{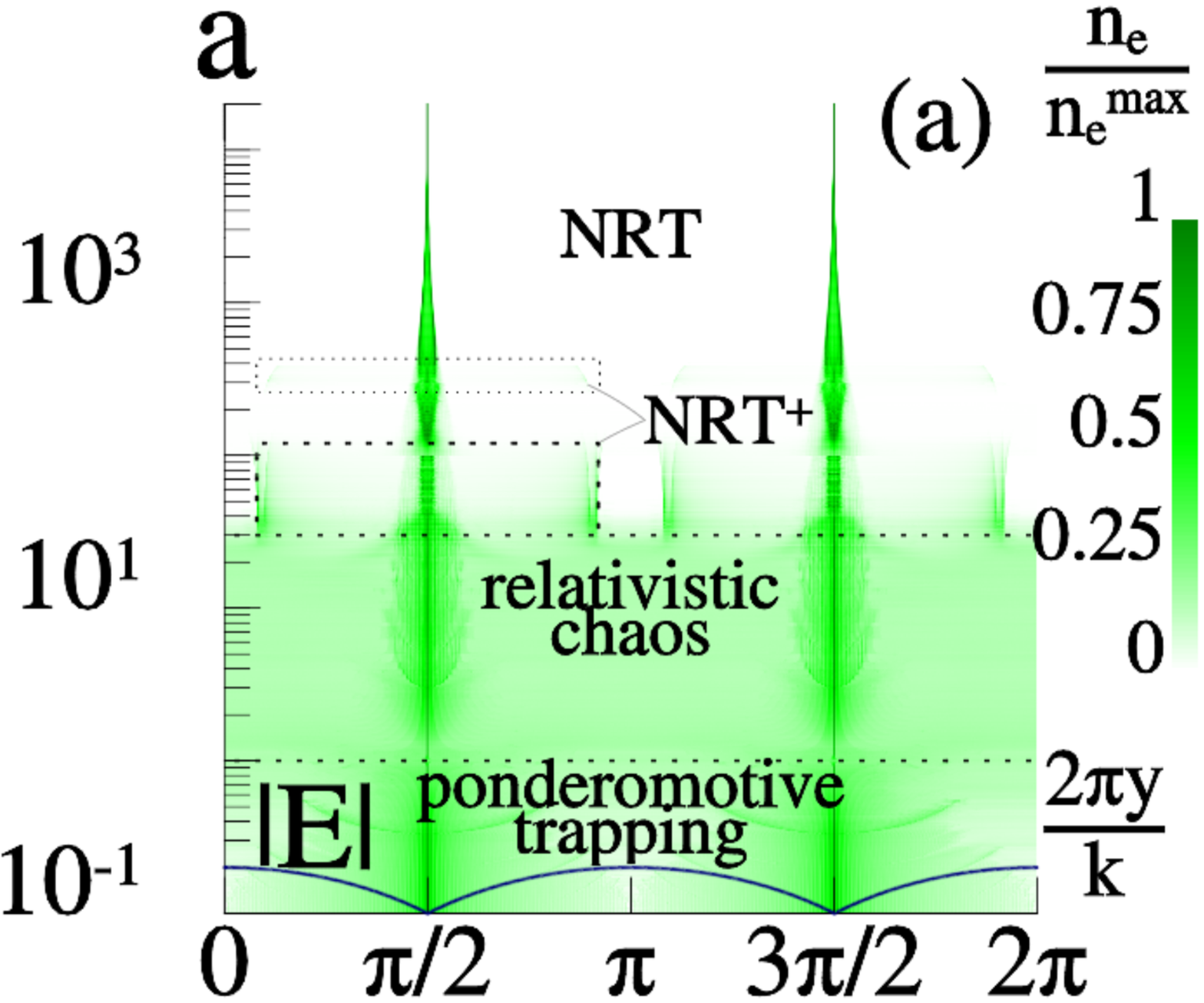}
		\includegraphics[width = 0.23\textwidth]{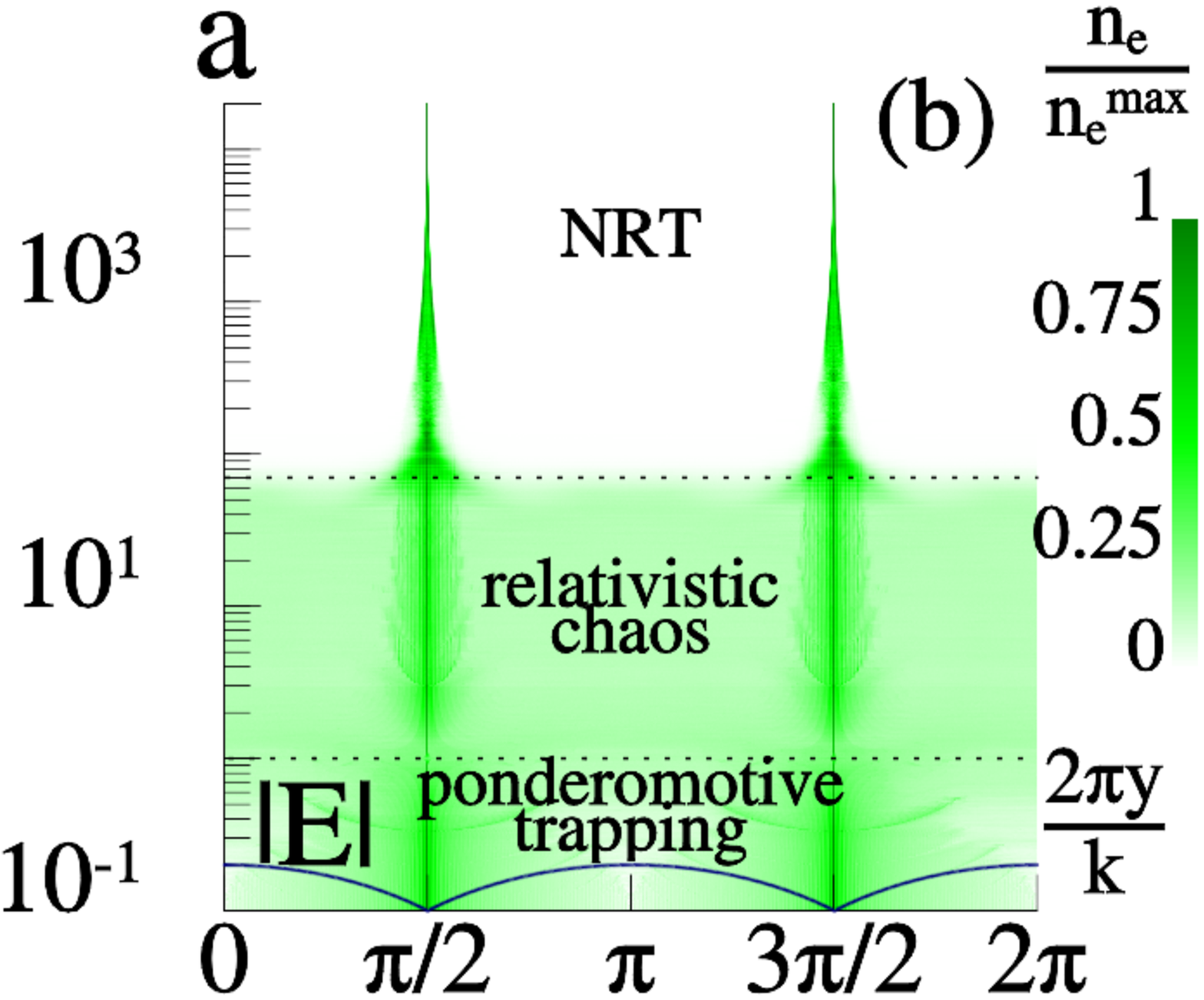}
		\includegraphics[width = 0.23\textwidth]{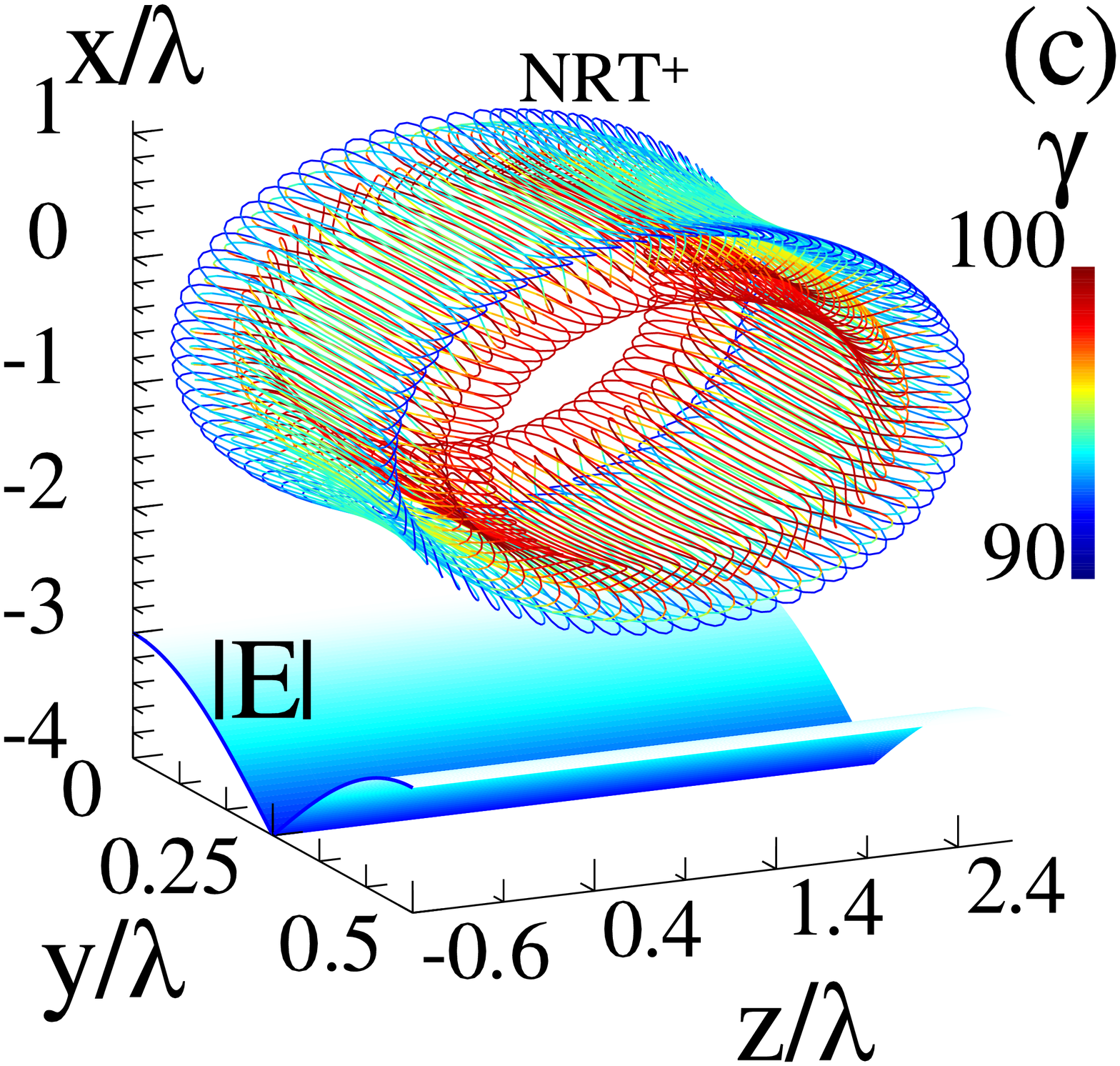}
		\includegraphics[width = 0.23\textwidth]{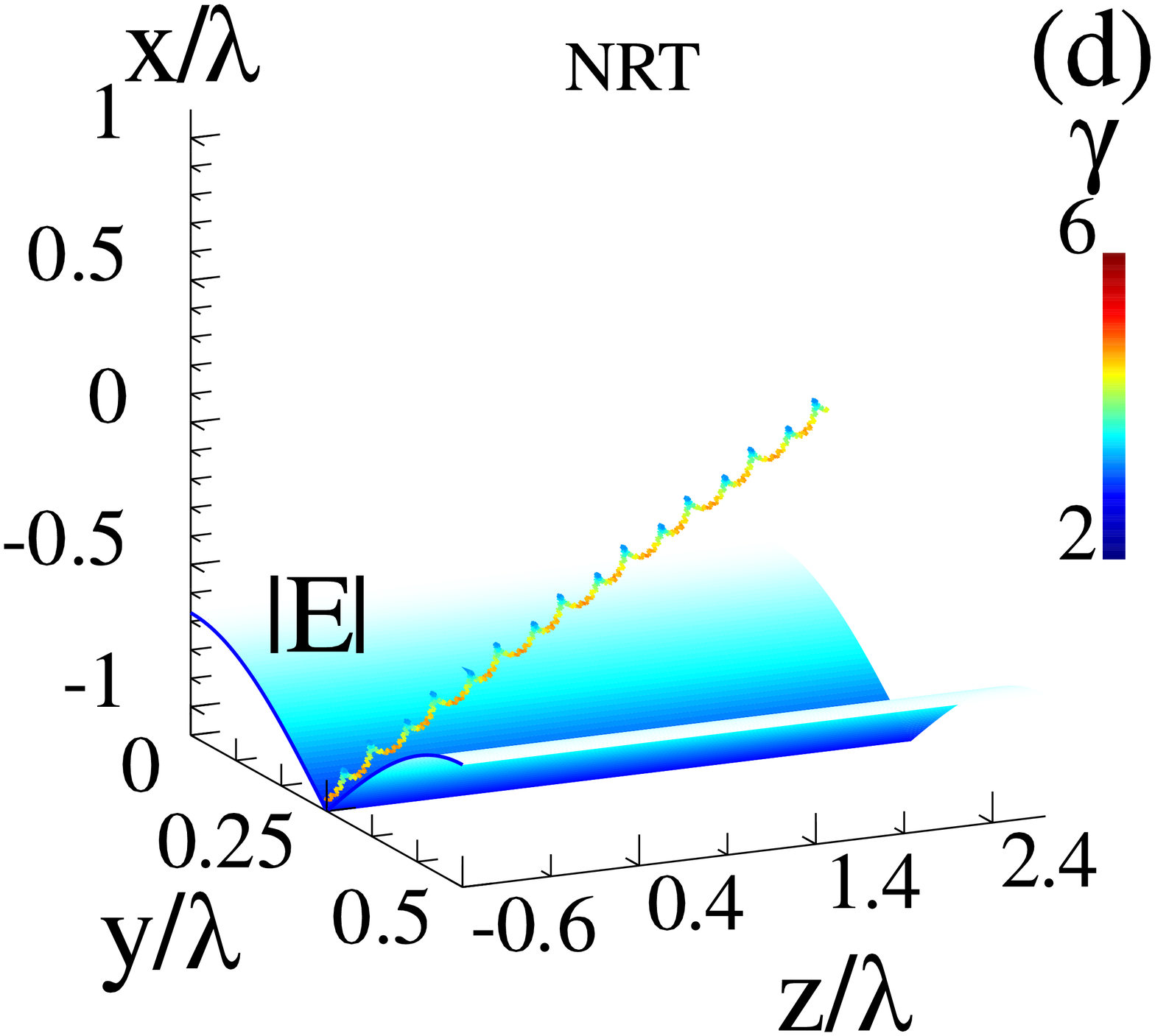}
	\caption{(Color online) Long term electron density distribution in the field of a plane circularly polarized standing wave as a function of wave amplitude in the frame of (a) approach (3) and (b) quasiclassical approach. NRT$^{+}$ (c) and NRT (d) trajectories for wave amplitude $a=100$. Color along the trajectory corresponds to gamma factor. Surfaces under the trajectories represent $|{\bf E}|$.}
	\label{mapQ}
\end{figure}

 
	Based on the introduced equations of motion it is possible to determine asymptotic regimes of motion in a circularly polarized field as was done in Ref. \cite{ART_Gonoskov} in a linearly polarized field. The results are shown in Fig.\ref{mapQ}. In the case of continuous force (Fig. \ref{mapQ}(a)) Eq. (\ref{FrrQ}) is used. Ponderomotive trapping, relativistic chaos and NRT can be revealed as in the case of a linearly polarized field. Relativistic effects at $a\gtrsim1$ lead to chaotization of motion. Particles do not accumulate at the electric field node, they can randomly pass from one node to another. Radiation reaction effects become apparent at smaller wave amplitudes $a_{\mathrm{NRT}}\approx30$ giving rise to NRT regime (for a linearly polarized wave $a_{\mathrm{NRT}}\approx400$). Although the radiation reaction force is much less than the Lorentz force at such amplitudes, over a long period of time the influence of dissipative force may be significant \cite{Tamburini_NuclInst,Lehmann_PRE}.\par
Along with the NRT trajectories there are special trajectories NRT$^+$ in the  $30<a<120$; $280<a<390$ amplitude ranges. Unlike the NRT trajectories localized in a small region around the electric field node, the amplitude of the oscillations along the $y$ axis in the NRT$^+$ regime is $\Delta y\approx0.4\lambda_l$, where  $\lambda_l$ is wavelength.  On these trajectories the electron goes to the region of a strong electric field and is reflected from them about every field period. The average energy is $\gamma_{NRT+}\approx a$, while in the NRT regime $\gamma_{NRT}\ll a$. Moreover, as follows from numerical simulations, the NRT$^+$ trajectory is localized, whereas in the NRT regime the particle is drifting in the transverse plane $xz$ with average velocity $0.6c$. The direction of drifting is determined by initial conditions. Examples of the trajectories are shown in Fig.\ref{mapQ}(c),(d).\par
There are also other special points of system (\ref{Plong})-(\ref{Yc}) $p_\|=0$, $y=\lambda_l n/4$ ($n\in \mathbb{Z}$), $p_\bot=0$. If a nonrelativistic electron appears in the vicinity of these points, it doesn't escape from this region \cite{Bashinov_PRE2015}. It is kept there by ponderomotive potential, which is in agreement with the vertical solid lines $y=\lambda_l/4,~3\lambda_l/4$ in Fig.\ref{mapQ}.\par
The use of Landau-Lifshitz force doesn't lead to qualitative changes in Fig.\ref{mapQ}(a). The only difference is a slight shift of the amplitude ranges of the NRT$^+$ regime to smaller amplitudes: $30<a<120$; $180<a<280$. Exclusion of radiation reaction stops formation of the NRT regime, and there is only relativistic chaos at relativistic amplitudes. The stochasticity of photon emission, on the contrary,  changes the motion regimes both quantitatively and qualitatively (Fig.\ref{mapQ}(b)) in the frame of the quasiclassical approach. NRT$^+$ regime doesn't arise,  which testifies to the continuity and discreteness of radiation losses.  The impact of photon emission on the electron doesn't allow the  NRT$^+$ regime to emerge and the electron can skip to the region of other electric field nodes, giving rise to relativistic chaos. Moreover, stochasticity of photon emission counteracts gradual cooling, which increases the threshold of NRT regime $a_\mathrm{NRT}=70$. It is interesting that there is no ART regime in a circularly polarized standing wave.

So, we have briefly described all stable asymptotic regimes, taking into account radiation losses. However, for fast processes like electron-positron pair production in extremely strong fields, the dynamic effects of motion can be very important. There is one more critical point at the electric field antinode.\par 

\section{The impact of radiation discreteness on electron motion: revision}
\label{Sec3}
It should be noted that the quantum-corrected RR force in Eqs.~(\ref{Plong})-(\ref{Yc}), used in most analytical treatments of ultrarelativistic particle dynamics, describes an average regular trajectory,  while due to radiation discreteness actual motion changes randomly at the instant of emission. This may affect average particle characteristics, such as mean relativistic factor (or mean energy) and rate of particle drifting. Such an impact was considered in Refs. \cite{Neitz_PRL,Bashinov_PRE2015,Yoffe_NJP2015} for a linearly polarized standing wave. In this section we address this issue to the circularly polarized wave, which, on the one hand, is a simpler field configuration but, on the other hand, a new effect of particle escape from a high-field region such as Brownian diffusion due to randomization of motion can also be generated.             

Apparently Eqs.~(\ref{Plong})-(\ref{Yc}) admit stationary trajectories in the plane of electric field antinode. They were studied earlier in Refs. \cite{Steiger_PRA1972,Zeldovich_UFN1975,Bulanov_PRE2011,Bashinov_PP2013,Zhang_NJP2015,Kostyukov_POP2016}. These trajectories are circles at the points $y=\pi n$ and are governed by the following equations
\begin{gather}
p_{\bot\mathrm{st}}=a\sin(\varphi_{\mathrm{st}}),\nonumber\\
a\cos(\varphi_{\mathrm{st}})=-F_{rst}p_{\bot\mathrm{st}},\nonumber\\
p_{\|\mathrm{st}}=0\label{ST},\\
\chi_\mathrm{st}=\eta\sqrt{a^2+p_{\bot\mathrm{st}}^4},\nonumber\\
\gamma_\mathrm{st}=\sqrt{1+p_{\bot\mathrm{st}}^2}=\sqrt{1+\frac{a^2}{1+F_{rst}^2}},\nonumber
\end{gather}
where $F_{r\mathrm{st}}$ is defined by Eqs.~(\ref{FLL}),(\ref{FrrQ}) with all variables replaced by the ones with subindex $\mathrm{st}$. Without radiation reaction we have $p_{\bot\mathrm{st}}=a$ and $\varphi_\mathrm{st}=\pi/2$. The angle between electron momentum and electric field $\varphi_\mathrm{st}$ becomes larger than $\pi/2$, due to radiation reaction and the electric field performs positive work compensating radiative losses. 

To make our statements clearer we briefly summarize the earlier studies. LL as well as LAD forces are known to overestimate radiation losses, consequently to underestimate energy, quantum parameter and to overestimate $\phi_{\mathrm{st}}$ at $a>a_Q\approx450$, which corresponds to $\chi\approx0.5$ at $\lambda_l=0.8\mu$m. Asymptotic behavior of the characteristics of the  trajectories can be represented in a simpler form. The radiation-dominated regime comes into force when $F_r\ge1$, the quantum regime demands $\chi\ge0.5$. If the radiation-dominated regime starts at $\chi\ll1$, then the asymptotic behavior of (\ref{FrrQ}) is the same as that of (\ref{FLL}), $F_{rst}\approx2\alpha\eta\gamma_\mathrm{st}^3/3\equiv\delta\gamma_\mathrm{st}^3$. As will be clear further, the validity condition $a\ll\frac{1}{\eta}$ \cite{Landau2} ensures $\gamma_\mathrm{st}^2>a$ and $\chi_\mathrm{st}\approx\eta\gamma_\mathrm{st}^2$. This case can be implemented if $\eta\ll\alpha^2/20$ , i.e. $\hbar\omega_l\ll1$eV. Otherwise, the radiation-dominated regime begins at larger amplitudes than the quantum regime. In the latter case, $F_{rst}=\frac{32\Gamma(2/3)}{3^{13/3}}\frac{\alpha\gamma_\mathrm{st}^{1/3}}{\eta^{1/3}}\equiv\delta_q\gamma_\mathrm{st}^{1/3}$ and $\eta$ should be $\eta\gg\alpha^2/10$. $\Gamma(x)$ is a gamma function. So, the optical frequency domain is at the boundary where the amplitude threshold of the 
radiation-dominated regime is very close to the quantum one. 

\begin{figure}[h]
	\centering
		\includegraphics[width = 0.15\textwidth]{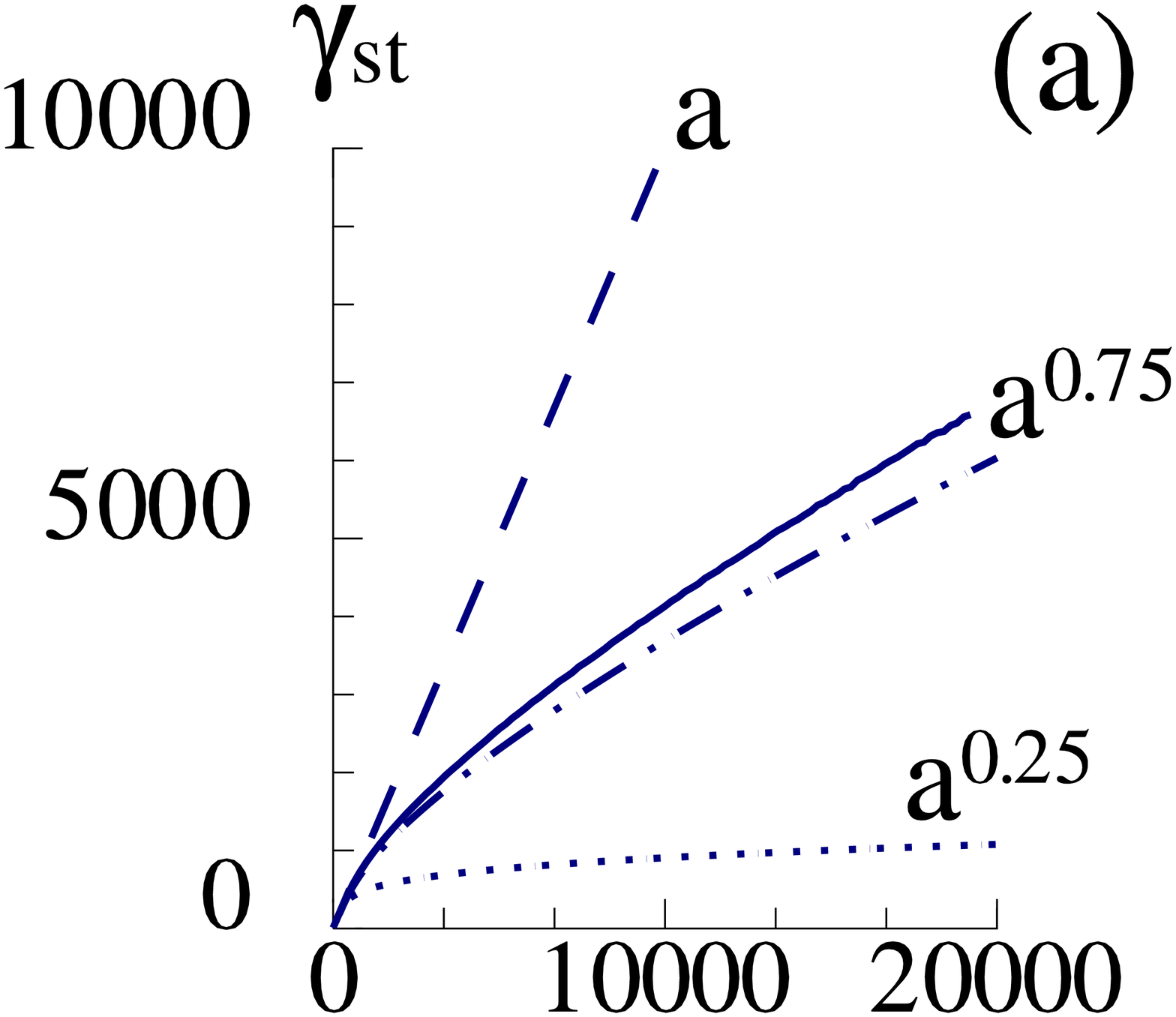}
		\includegraphics[width = 0.15\textwidth]{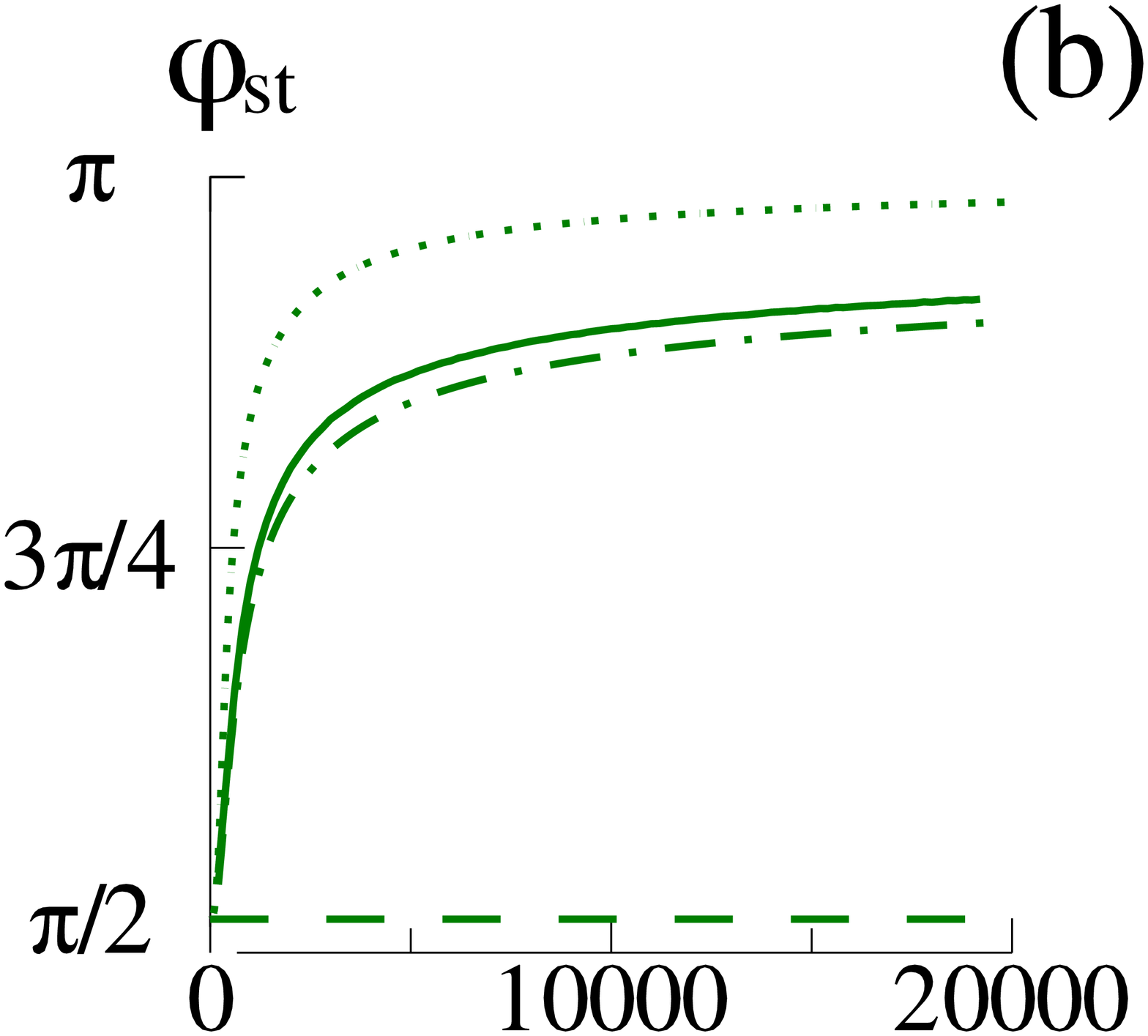}
		\includegraphics[width = 0.15\textwidth]{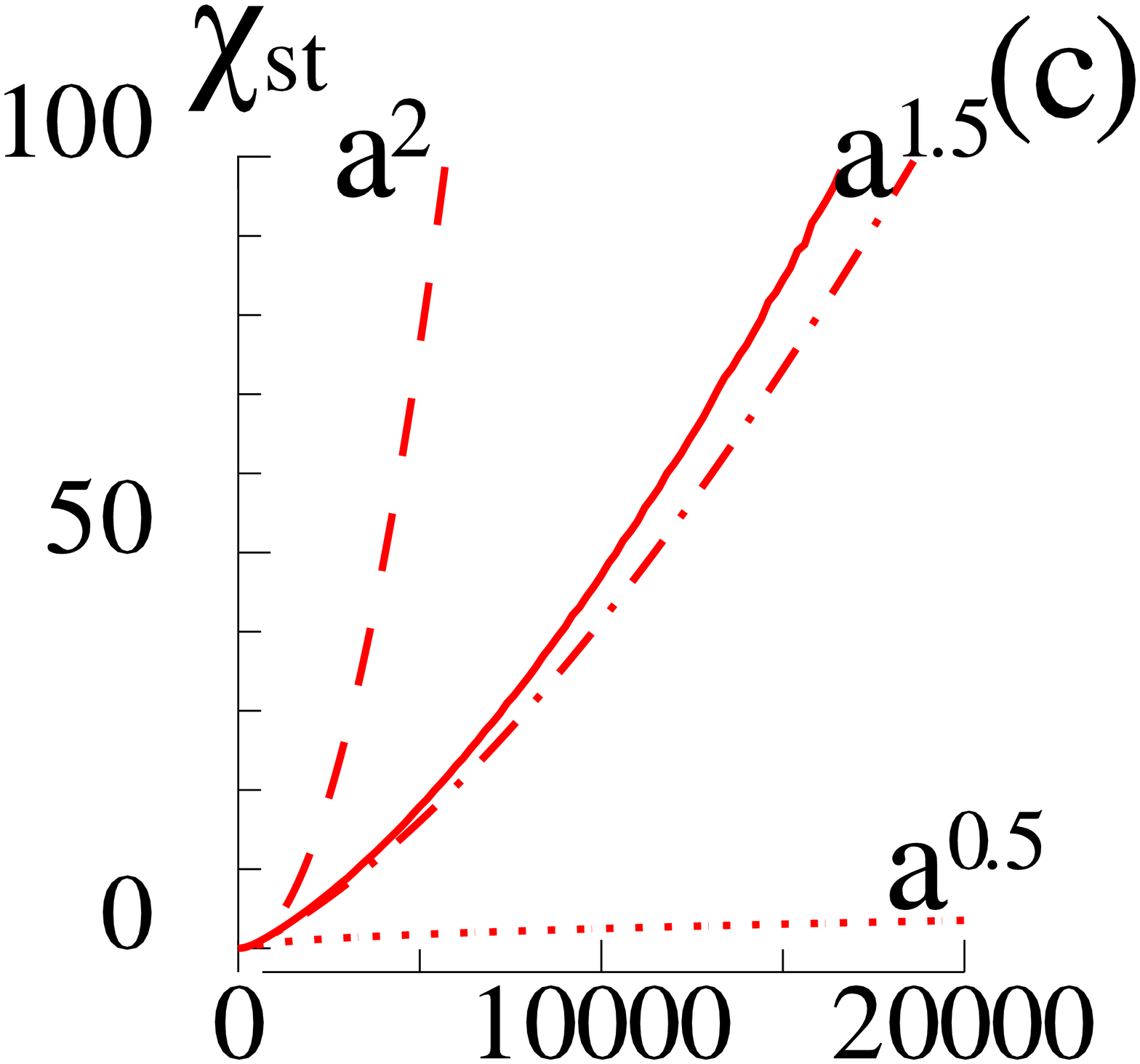}
		\caption{ (Color online) (a) Lorentz factor, (b) angle between electric field and electron momentum, (c) quantum parameter $\chi_\mathrm{st}$ at stationary trajectory as a function of the field amplitude of standing circularly polarized wave. The dash-dotted and dotted lines correspond to allowance for radiation reaction force with quantum corrections (\ref{FrrQ}) and in accordance with Landau-Lifshitz equation (\ref{FLL}), respectively. Values obtained without radiation reaction are depicted by the dashed line. The solid line corresponds to quasiclassical case. Formulas show asymptotic behavior of trajectory parameters.}
		\label{fig:StGAH}
\end{figure}

Next, we focus on the optical frequency domain where the most powerful laser sources are expected. For the wavelength $\lambda_l=0.8\mu$m, thresholds of radiation-dominated and quantum regimes are $a_{RR}=(2\alpha\eta/3)^{-1/3}\approx400$ and $a_Q=0.2\alpha/\eta\approx450$, respectively, so $a_{RR}\approx a_{Q}$. To generalize the expression for the gamma factor in the radiation-dominated or quantum regime we introduce 
\begin{equation}
\gamma_\mathrm{st}=\left(\frac{a}{\tilde{\delta}}\right)^{1-s},
\label{GLLQ}
\end{equation}
where for LL or LAD forces $\tilde{\delta}=\delta,~~s=3/4$ and for quantum corrected force $\tilde{\delta}=\delta_q,~~s=1/4$. Using this asymptotic behavior and assuming $a<1/\eta$ we can simply find that $\gamma_\mathrm{st}^2>a$. The radiation reaction force and the quantum parameter are
\begin{gather}
F_{r\mathrm{st}}=Da^{s}\label{FrrG},\\
\chi_\mathrm{st}=\eta D^{-2}a^{2(1-s)}\label{ChiG},
\end{gather}
where $D=\tilde{\delta}^{1-s}$.  Without radiation reaction, $\gamma_\mathrm{st}=a$ and $\chi_\mathrm{st}=\eta a^2$. Thus, characteristics of the trajectories are very sensitive to the way we describe radiation losses.  It is worth noting that for the quantum-corrected RR force, Eqs.~(\ref{GLLQ}),(\ref{ChiG}) approximate the gamma factor $\gamma_\mathrm{st}^\mathrm{corr}$ and quantum parameter $\chi_\mathrm{st}^\mathrm{corr}$ well at wave amplitudes $a\sim100000$, while at reasonable values of $a$ of the order of several thousands and $\lambda_l\approx0.8\mu$m, more suitable approximation is $\gamma_\mathrm{st}^\mathrm{corr}\approx6.17a^{0.695}$ and $\chi_\mathrm{st}^\mathrm{corr}\approx1.8\cdot10^{-4}a^{1.34}$. The energy $\gamma_{\mathrm{st}}$, quantum parameter $\chi_{\mathrm{st}}$ and angle $\phi_{\mathrm{st}}$ characterizing stationary trajectories with different description of RR forces are compared in Fig.~\ref{fig:StGAH} (see also, e.g., \cite{Zhang_NJP2015}). 
However, the quantum nature of photon emission is especially important at $\chi\gtrsim1$: electrons randomly emit hard photons losing a significant part of their momentum and energy and after that they are accelerated fast again. In this case, due to random photon emission the particle motion is irregular, it rather has a fragmentary nature. Therefore, the averaged gamma factor as well as the quantum parameter don't need to have the same values as in the case with the quantum-corrected RR force. In the quasiclassical case, the particles have a possibility to gain energy up to $a$,  while  in the radiation-dominated regime the mean energy should be proportional to $a^{0.75}$, and the mean value of $\chi$ proportional to $a^{1.5}$ \cite{Fedotov_PRL2010}. For determining the dependence of $\gamma_\mathrm{st}^\mathrm{Q},\chi_\mathrm{st}^\mathrm{Q},\varphi_\mathrm{st}^\mathrm{Q}$ on $a$ for the case of interest we calculate the motion of 1000 particles at the antinode of electric field taking into account the stochasticity of emission. In the long-term evolution of such an ensemble, when the distribution function in momentum space is stabilized, we determine the corresponding mean values. So, according to the numerical simulations the gamma factor $\gamma_\mathrm{st}^\mathrm{Q}$ and quantum parameter $\chi_\mathrm{st}^\mathrm{Q}$ are greater than the corresponding values of $\gamma_\mathrm{st}^\mathrm{corr},\chi_\mathrm{st}^\mathrm{corr}$ in the case of quantum-corrected RR force, and the angle $\varphi_\mathrm{st}^\mathrm{Q}$ is closer to $\pi$. These parameters as functions of $a$ are varied as follows: $\gamma_\mathrm{st}^\mathrm{Q}=1.13\gamma_\mathrm{st}^\mathrm{corr}$ and $\chi_\mathrm{st}^\mathrm{Q}=1.15\chi_\mathrm{st}^\mathrm{corr}$. 
As we see, the stochasticity of photon emission just slightly corrects the mean values of $\chi$ and $\gamma$ but  more importantly it generates the new effect of particle diffusion mainly in the transverse directions. This occurs because each act of photon emission breaks the invariant ${\bf p_\bot}-{\bf A}={\bf const}$, which causes additional drifting in the direction opposite to the photon momentum. Note that this diffusion exists even in a plane wave as it is connected with the stochastic nature of photon emission. A relativistic particle is shifted along a certain direction from its initial position after emission of $n$ photons roughly at $\Delta=\sum_{i=0}^{n}\cos(\psi_i)c/W_\gamma$. $\psi$ is the angle between the direction and drift velocity after photon emission, and $W_\gamma$ is the probability of photon emission per unit time, $W_\gamma\approx\frac{1.46\alpha\omega_l\chi^{2/3}}{\eta\gamma}$ in the case of $\chi\gg1$ \cite{BayerKatkov}. $\psi_i$ is uniformly distributed in the $0..2\pi$ range. Thus, the mean value is $\mu\Delta=0$, dispersion is $\Theta\Delta=0.5nc^2/W_\gamma^2$, and diffusion coefficient is $d=0.25c^2/W_\gamma$ or in dimensionless variables 
\begin{equation}
d\approx\frac{\eta^{1/3}}{6\alpha{\gamma_\mathrm{st}^\mathrm{Q}}^{1/3}}.
\label{Diff}
\end{equation}

\section{Longitudinal drifting}

As follows from the long-term density distributions shown in Fig.~\ref{mapQ}, the particles mainly tend to move from the high electric field (antinode) region to the minimum ponderomotive potential (node region). This is a quite expected result for a standing wave with circular polarization, although we would like to note that the NRT$^{+}$ regime (Fig.~\ref{mapQ}(a,c)) with classical description of RR force and as well the ART regime in a linearly polarized standing wave \cite{ART_Gonoskov} were unexpectedly new. However, for the problem of interest when pair plasma is generated primarily in the antinode region, it is important to know the rate of particle escape. To do so, we study the influence of radiation on the stability of the stationary trajectory given by Eqs. (\ref{ST}). Assuming  $\mathbf{p}=\mathbf{p_{\mathrm{st}}}+\tilde{\mathbf{p}}(t)$, $\varphi=\varphi_{\mathrm{st}}+\tilde{\varphi}(t)$, $y=y_{\mathrm{st}}+\tilde{y}(t)$, where $\tilde{\mathbf{p}}(t)$, $\tilde{\varphi}(t)$, $\tilde{y}(t)$ are small perturbations, we substitute them into Eqs.~(\ref{Plong}) - (\ref{Yc}). Then keeping only linear terms of the perturbations the governing equations are written as
\begin{gather}
\frac{d\tilde{p_\bot}}{dt}=p_{\bot\mathrm{st}}\tilde{\varphi}-\tilde{F_r}p_{\bot\mathrm{st}}-\tilde{p_\bot}F_{r\mathrm{st}}, \label{tPtr}\\
\frac{d\tilde{\varphi}}{dt}=\frac{1}{p_{\bot\mathrm{st}}}(-\tilde{p_\bot}+a\cos(\varphi_{\mathrm{st}})\tilde{\varphi}),\label{tPhi}\\
\frac{d\tilde{p_{\|}}}{dt}=\frac{p_{\bot\mathrm{st}}^2\tilde{y}}{\gamma_{\mathrm{st}}}-F_{r\mathrm{st}}\tilde{p_\|},\label{tPlong}\\
\frac{d\tilde{y}}{dt}=\frac{\tilde{p_\|}}{\gamma_{\mathrm{st}}}\label{tYc}.
\end{gather}
Without loss of generality we assume $y_{\mathrm{st}}=0$. For comparative analysis we consider again different descriptions of radiation losses.\\
1) LL force:
\begin{equation}
\tilde{F_r}=-\frac{2\alpha\eta}{3\gamma_{\mathrm{st}}}a^2p_{\bot\mathrm{st}}^2\sin(2\varphi_{\mathrm{st}})\tilde{\varphi}+\frac{2\alpha\eta\tilde{p_\bot}p_{\bot\mathrm{st}}}{3\gamma_{\mathrm{st}}^3}\left(\gamma_{\mathrm{st}}^4-1-a^2\right).
\label{tFLL}
\end{equation}
2) Quantum-corrected case:
\begin{equation}
\begin{split}
\tilde{F_r}=&\frac{\alpha p_{\bot\mathrm{st}}}{3\sqrt{3}\pi\eta\gamma_{\mathrm{st}}}\left[\frac{p_{\bot\mathrm{st}}^2\left(\tilde{p_\bot}+a\cos(\varphi_{st})\tilde{\varphi}\right)}{3\chi_{\mathrm{st}}\left(a^2+p_{\bot\mathrm{st}}^4\right)}\right.\\
\int_0^\infty& \frac{u^2(4u^2+5u+4)}{(1+u)^4}\left(K_{1/3}\left(\frac{2u}{3\chi_\mathrm{st}}\right)+K_{5/3}\left(\frac{2u}{3\chi_\mathrm{st}}\right)\right)du-\\
\frac{\tilde{p_\bot}}{\gamma_\mathrm{st}^2}&\left.\int_0^\infty \frac{u(4u^2+5u+4)}{(1+u)^4}K_{2/3}\left(\frac{2u}{3\chi_\mathrm{st}}\right)du \right].
\end{split}
\label{tFrrQ}
\end{equation}
The system ~(\ref{tPtr})-(\ref{tYc}) can be divided into two pairs of equations (\ref{tPlong}), (\ref{tYc}) and (\ref{tPtr}), (\ref{tPhi}). Solutions  are written in the form $\tilde{p_\|},\tilde{y}\propto e^{\lambda_\| t}$ and $\tilde{p_\bot},\tilde{\varphi}\propto e^{\lambda_\bot t}$. Without radiation reaction $\lambda_{\|\pm}=\pm\frac{\sqrt{\gamma^2-1}}{\gamma}$, in the ultrarelativistic case $\lambda_{\|\pm}=\pm1$ and $\lambda_\bot=0$. In a general case 
\begin{equation}
\lambda_{\|\pm}=0.5\left(-F_{r\mathrm{st}}\pm\sqrt{F_{r\mathrm{st}}^2+4p_{\bot\mathrm{st}}^2/\gamma_\mathrm{st}^2}\right) .\label{Llong}
\end{equation}
This means that $y=\pi n/2$ is a saddle point, and the positive value corresponds to the rate of particle drifting along the $y$ axis to the electric field node. The asymptotic behavior of $\lambda_{\|\pm}$ for $a>a_{RR}$ is
\begin{gather}
\lambda_{\|-}\approx -F_r,\nonumber\\
\lambda_{\|+}\approx 1/F_r.\label{Llapp}
\end{gather}
In the case of LL force (as well as LAD force) $\lambda_{\|+}^{\mathrm{LL}}\propto a^{-0.75}$, for quantum-corrected force $\lambda_{\|+}^{\mathrm{corr}}\propto a^{-0.25}$. So, quantum corrections change not only the factor of power function, but also the power law, and modify significantly the rate of longitudinal drifting. Note that in the case of quantum-corrected force, $\lambda_{\|\pm}^{\mathrm{corr}}$ approaches the asymptotic behavior at $a\sim100000$, whereas for the considered parameters $\lambda_{\|+}^{\mathrm{corr}}\approx5.5a^{-0.3}$.

\begin{figure}[h]
	\centering
		\includegraphics[width = 0.15\textwidth]{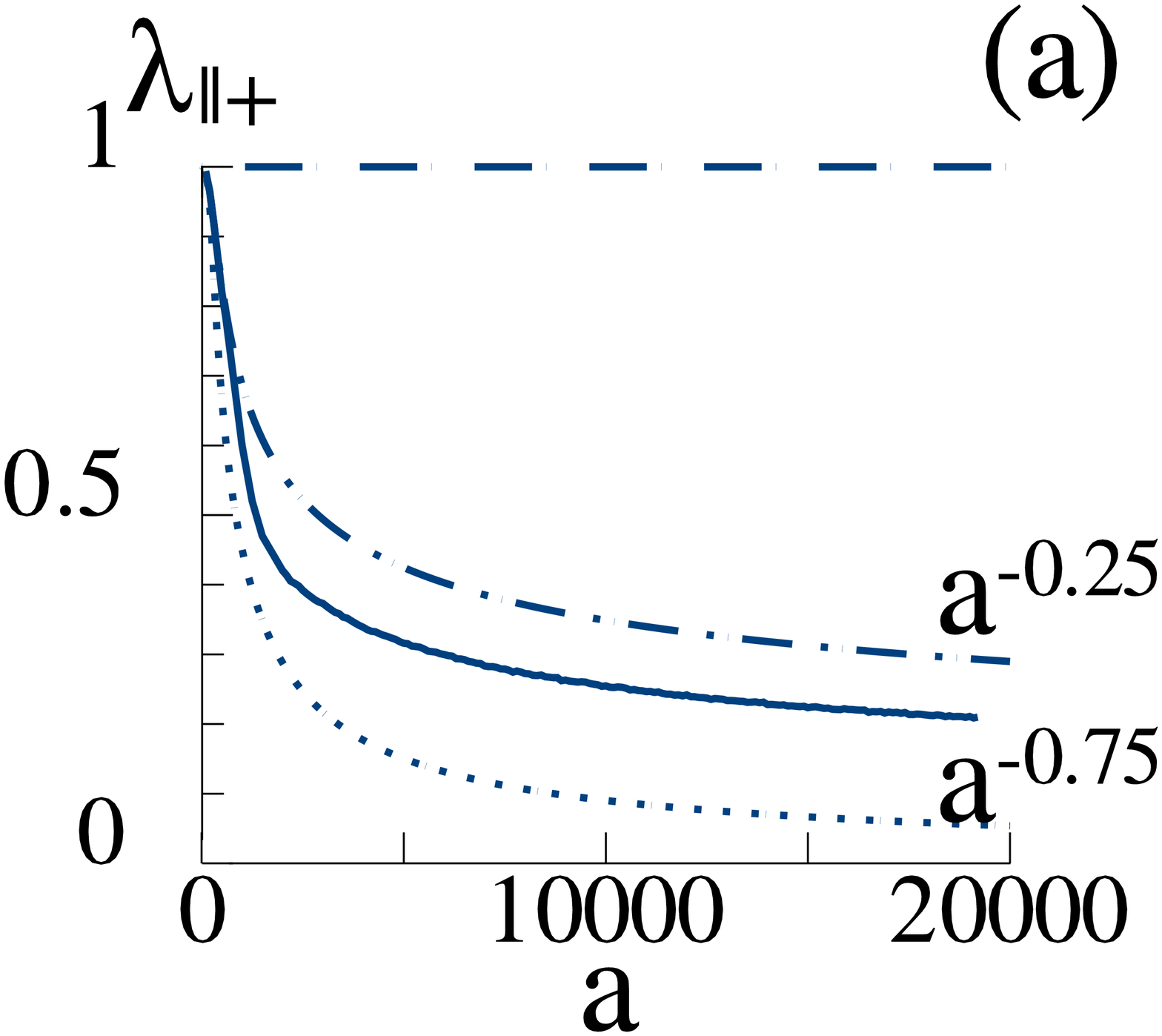}
		\includegraphics[width = 0.15\textwidth]{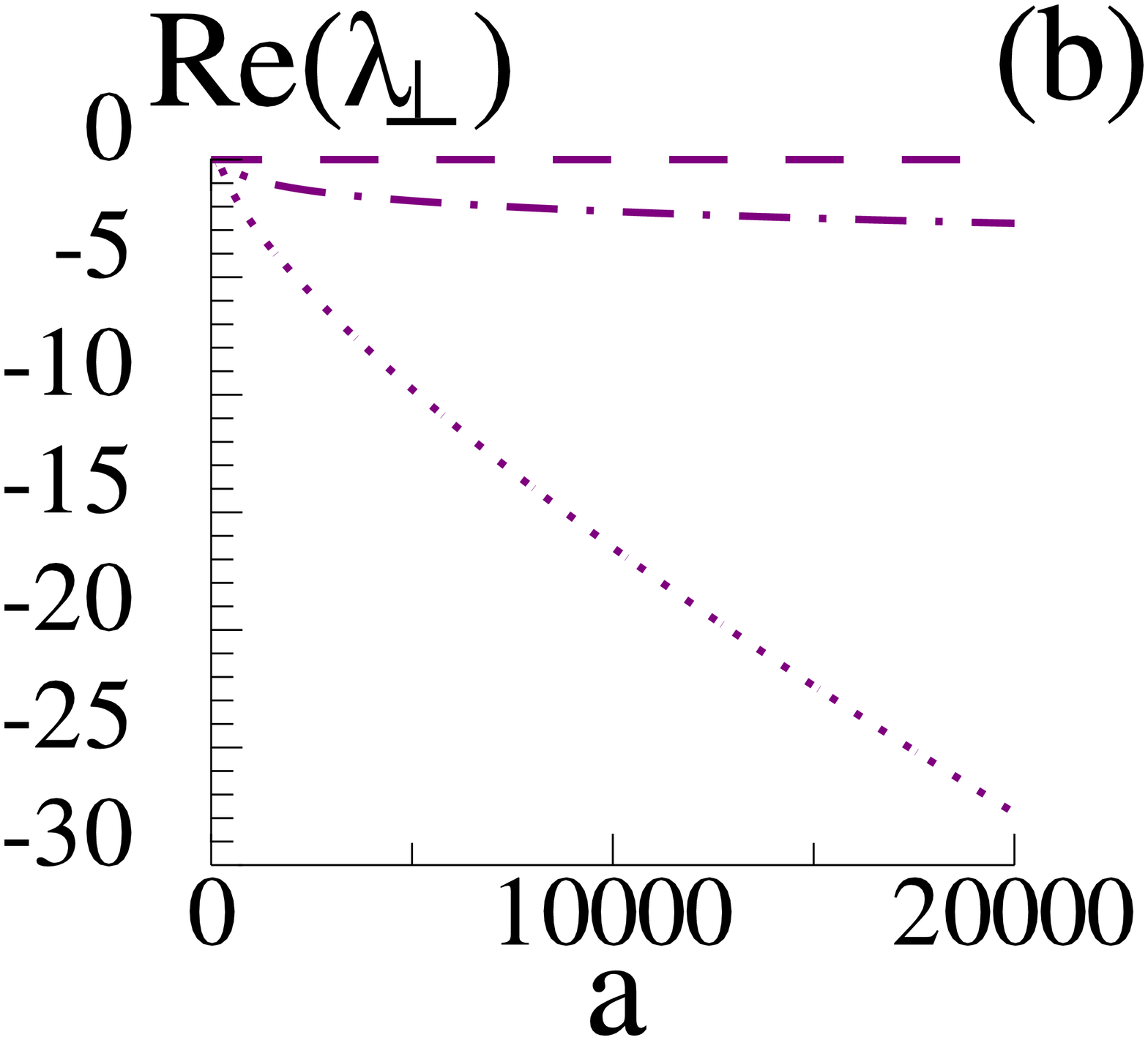}
		\includegraphics[width = 0.15\textwidth]{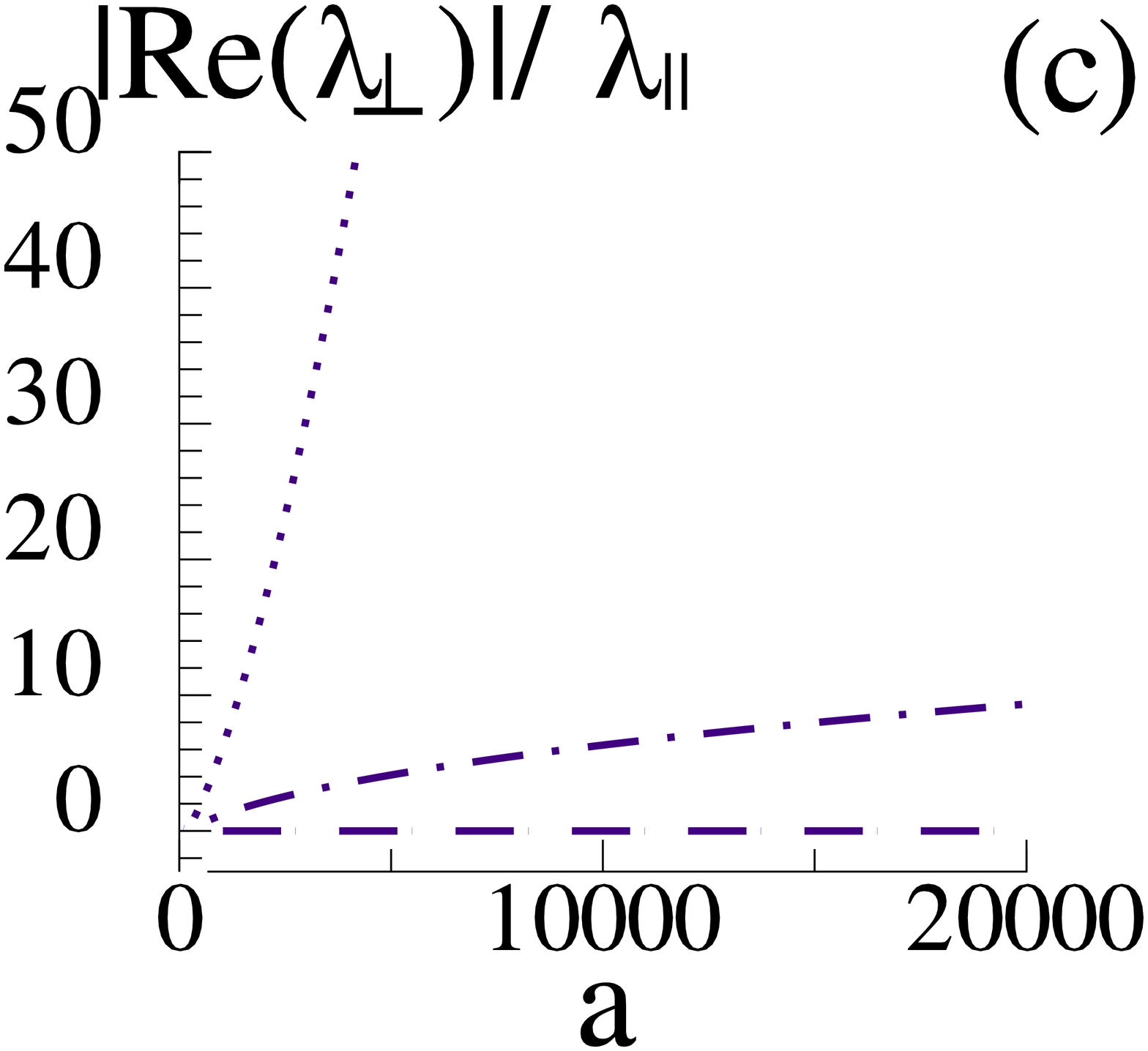}
		\caption{(Color online) (a) $\lambda_{\|+}$ escape rate of electrons along the $y$ axis, (b) $\mathrm{Re}(\lambda_\bot)$ rate of approach to the stationary trajectory in transverse direction and (c) $|\mathrm{Re}(\lambda_\bot)|/\lambda_{\|+}$ ratio as a function of field amplitude of standing circularly polarized wave. Line style is the same as in Fig.\ref{fig:StGAH}}
		\label{fig:Rates}
\end{figure}

The two roots of $\lambda_\bot$ are complex conjugates having negative real parts. The stable focus is in the phase plane $\tilde{p_\bot},~~\tilde{\phi}$. $\lambda_\|$ and $\lambda_\bot$ can be found numerically using Eqs. (\ref{tPtr})-(\ref{tYc}). $\lambda_{\|+}$, ${\mathrm Re}\lambda_\bot$ and their ratio are shown in Fig.\ref{fig:Rates}.
First, the radiation reaction slows down the longitudinal drift and accelerates approach to the stationary trajectory in the transverse plane ($p_\bot,\phi$). In the ultrarelativistic case, electrons emit photons in a cone with a small angle around electron velocity, thus the photon emission counteracts motion in the direction of velocity. The stronger the radiation losses,  the slower the longitudinal drift is. At the same time, in the transverse direction the particles tend to the stationary trajectory, where losses are compensated by positive work of the field. Second, when $a>1500$ (in the case of classical approach with quantum corrections), we have the $|\mathrm{Re}(\lambda_\bot^\mathrm{corr})|/\lambda_{\|+}^\mathrm{corr}>1$ ratio  and the particles first quickly approach the stationary trajectory and after that slowly drift in the longitudinal direction to the electric field node. So, the trajectory can be characterized by local values of the field, and inertia of particle motion can be neglected. Radiation reaction retards particles in the vicinity of the electric field antinode, and the characteristic time of the  drift is $t_{n}\propto\lambda_{\|+}^{-1}$. The fact that $t_{n}$ can be much longer than the optical cycle is explained not only by initial proximity to the electric field antinode, but also by  the small value of $\lambda_{\|+}$, which is important. In the case of using the LL (LAD) force, $\lambda_{\|+}$ and ${\mathrm Re}\lambda_\bot$ are underestimated at $a>a_{Q}$.

The qualitative behavior of $\lambda_{\|+}$ can be characterized by considering longitudinal motion in the radiation-dominated regime. In this regime $p_\|\ll p_\bot$, so the characteristics of the trajectory are determined by the local value of the field $a(y)=a\cos y$. Then the exact expression for ponderomotive force from Eqs.(\ref{Plong}), (\ref{ST}) is
\begin{equation}
F_p=-\frac{\left[a^2(y)\right]'}{2\gamma(1+F_r^2)}=\frac{a^2\sin(2y)}{2\gamma(1+F_r^2)}.
\label{PondForce}
\end{equation}
The same expression for ponderomotive force in the case of LAD force has been obtained in Ref.\cite{Bashinov_PP2013}. Then assuming $\partial p_\|\partial t\simeq 0$ in Eq.(\ref{Plong}) we can obtain expressions for longitudinal momentum and velocity in the radiation-dominated regime as long as $|{\bf E}(y)|\gg|{\bf B}(y)|$:
\begin{gather}
p_\|=\frac{F_p}{F_r}=\frac{\sin(2y)}{2D^2a^{2s-1}\cos^{2s+1}(y)},\label{py}\\
v_\|=\frac{F_p}{\gamma F_r}=\frac{\sin(2y)}{2Da^s\cos^{2+s}(y)} \label{vy}.
\end{gather}
The time of drifting from the antinode to the node of the electric field $t_n=\int dy/v_\|$, and $\lambda_{\|+}\propto a^{-s}$ is in agreement with (\ref{Llapp}).

In order to compare longitudinal drifting within the classical and quasiclassical approaches, we will consider the evolution of $N_e=1000$ electrons initially at rest located at the point $y=0.001\lambda_l$ and determine the period of time $t_n$ when the electrons center mass reaches the point $y=\lambda_l/4$. In the quasiclassical case, electron evolution is drift and diffusion. In the classical case, when $(\mathrm{Re}\lambda_\bot)/\lambda_{\|+}\gg1$, it follows from Eqs. (\ref{tPtr})-(\ref{tYc}) that the trajectory is described by the expression
\begin{equation}
y=\frac{y_0}{\lambda_{\|-}-\lambda_{\|+}}\left(\lambda_{\|-}e^{\lambda_{\|+}t}-\lambda_{\|+}e^{\lambda_{\|-}t}\right).
\label{Ylong}
\end{equation}
\begin{figure}[h]
	\centering
		\includegraphics[width= 0.2\textwidth]{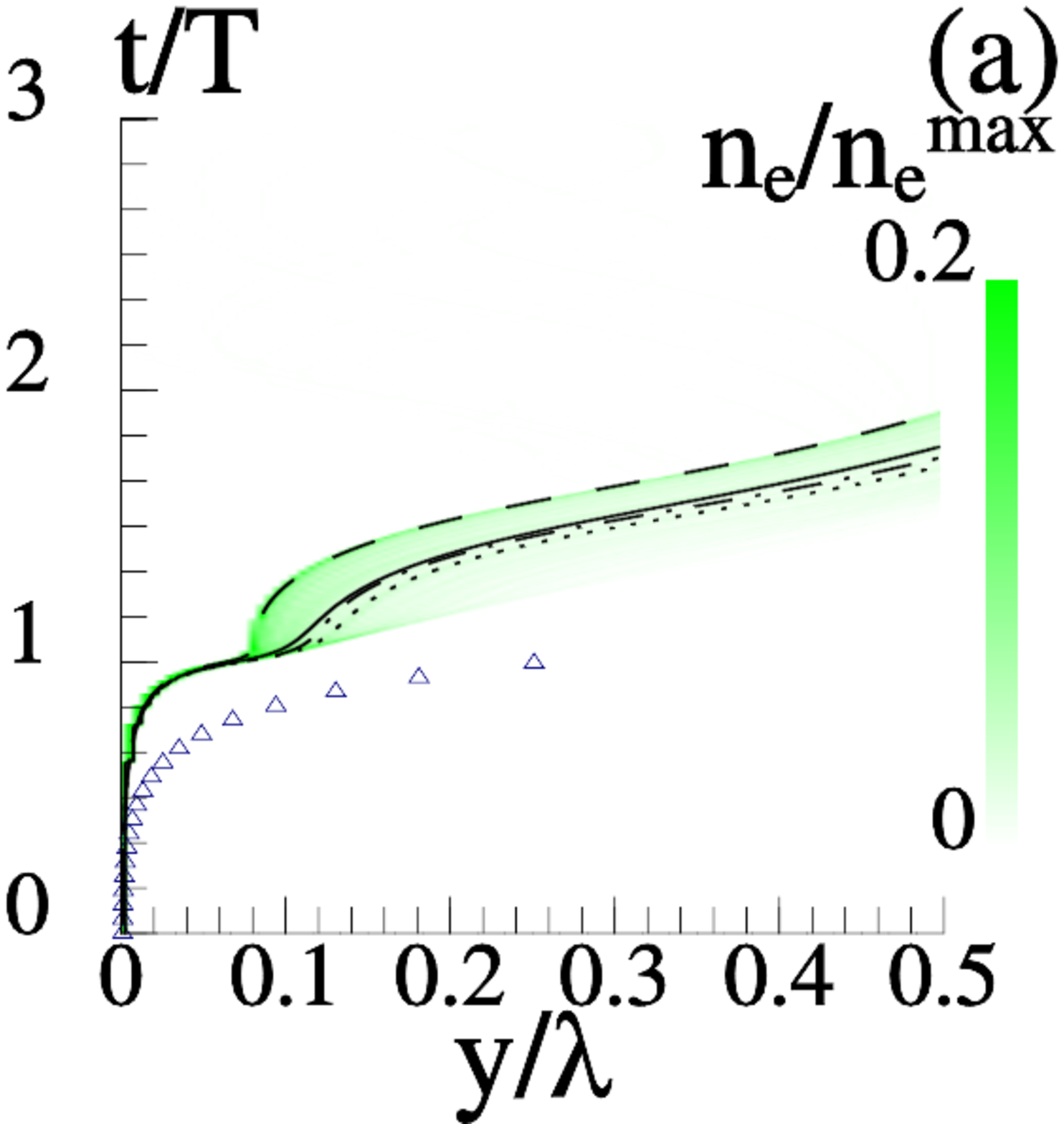}
		\includegraphics[width= 0.2\textwidth]{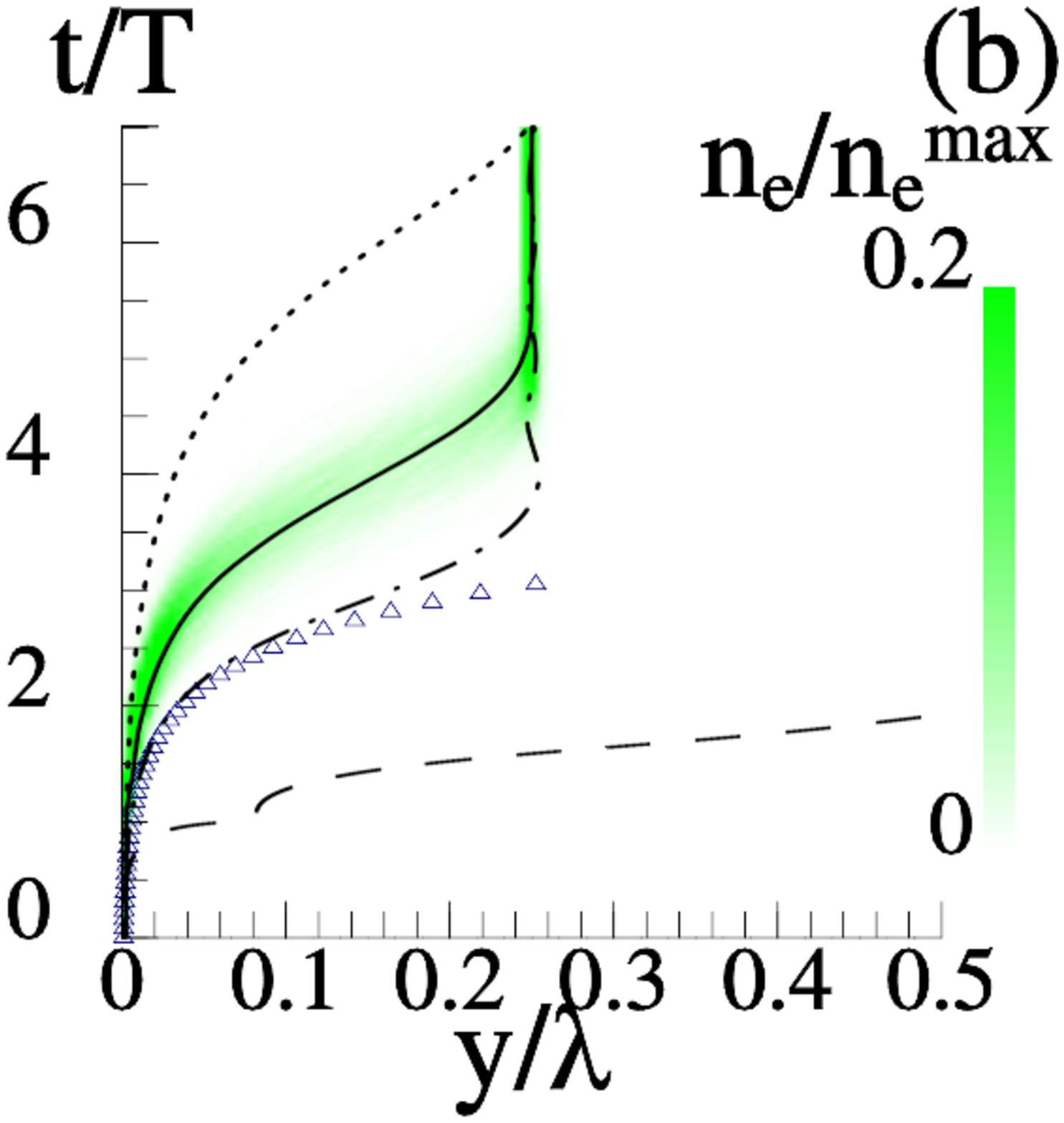}
	\caption{(Color online) Evolution of electron distribution along $y$ axis for wave amplitude (a) $a=100$, and (b) $a=20000$ in the quasiclassical case. Solid line corresponds to trajectory of mass center. Trajectories represented by dashed and dotted lines were obtained without and with radiation reaction in the form of Landau-Lifshitz taken into account, respectively. Quantum corrections to Landau-Lifshitz force give the trajectory depicted by the dash-dotted line. The line marked by triangles corresponds to Eq.(\ref{Ylong}). Time along the dotted line in Fig. (b) is 2.5 times faster.}
	\label{fig:YZ}
\end{figure}

For small amplitudes $a<a_{RR}$ the influence of radiation losses over a short period of time is weak. Figure~\ref{fig:YZ}(a) corresponds to $a=100$. The trajectories obtained with and without radiation reaction are approximately the same as long as the particles don't reach the region of strong magnetic field. After that, diffusion due to stochasticity of photon emission smooths the electron distribution. The upper boundary of electron distribution corresponds to the electrons that don't have enough time to emit a photon (dashed line in Fig.\ref{fig:YZ}(a)). Motion of the electrons center of mass can be described within the framework of the  classical approaches to radiation reaction description (solid, dotted and dash-dotted lines in Fig.\ref{fig:YZ}(a)). However, the trajectory considered above under certain initial conditions does not approach the described stationary trajectory (the line marked by triangles differs from, for example, the dotted line in Fig.\ref{fig:YZ}(a)). In this case a dynamical stop effect occurs  at $y\approx0.1\lambda_l$. Radiation losses smooth this effect, for the same reason the slowest motion occurs without photon emission. Particles are not trapped at the closest node region.

In the case of large amplitudes $a\gg a_{RR}$, radiation losses change particle motion qualitatively. Figure \ref{fig:YZ}(b) corresponds to the extremely strong field with amplitude $a=20000$. First, particles are trapped by the region of the closest electric field node. Second, initial conditions can be neglected, that's why there is no stop effect, if radiation reaction is taken into account. Particles approach the stationary trajectory and drift slowly to the electric field node. This is clear from comparison of the dash-dotted curve (radiation reaction (\ref{FrrQ}) is taken into account) and the curve corresponding to (\ref{Ylong}) marked by triangles. The difference between the curves arises in the region of weak electric and strong magnetic field, where (\ref{Ylong}) is not valid. The slowest longitudinal drift is typical for particles that have experienced the greatest radiation losses. Without radiation reaction, the trajectory is approximately the same as that for $a=100$. In other approaches, the drift is essentially slower in the case of LL (LAD) force or excessively fast in the case of quantum corrections (\ref{FrrQ}). In fact, the use of continuous radiation reaction force is not applicable when the quantum parameter of the particle $\chi\gtrsim1$. In this case, the particle can lose a substantial part of its energy, and consequently the same part of longitudinal momentum. It needs additional time to be accelerated, to approach the stationary trajectory and to obtain longitudinal momentum. Recently it was noticed on an example of trajectories that, on the average, particles drift longer to the electric field node in the quasicalssical case than in the case when radiation losses are described as continuous force \cite{Duclous_PPCF2011}. That phenomenon was explained by the straggling effect. The reason of the difference can be clearer from comparison  of ponderomotive forces (proportional to $\sin(\varphi)$ as follows from Eq.(\ref{Plong}), $\varphi$ is shown in Fig.\ref{fig:StGAH}(b)). In the quasiclassical case $\sin(\varphi)$ is  less than in the case of quantum-corrected force. Thus, the  quasiclassical approach gives a more correct result taking into account energy losses and stochasticity of photon emission.

\begin{figure}[h]
	\centering
		\includegraphics[width = 0.25\textwidth, height = 0.25\textwidth]{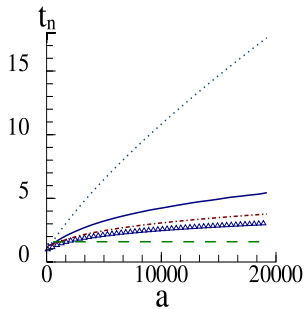}
	\caption{(Color online) Time $t_n$ needed for the electron to reach the electric field node as a function of wave amplitude. The solid line corresponds to quasiclassical approach. Dotted and dash-dotted lines obtained taking into account radiation reaction force in the form of Landau-Lifshitz and with quantum corrections respectively. The dashed line represents $t_n$ without radiation losses. Line marked by triangles obtained from numerical solution of Eq.~(\ref{Ylong}) with $y$ substituted for $\pi/2$.}
	\label{fig:tn}
\end{figure}
The time spent by the particle to reach the electric field node $t_n$ as a function of wave amplitude is shown in Fig.\ref{fig:tn}. Without radiation reaction, $t_n$ is approximately constant $t_n\approx1.59$. In a general case, as a result of radiation losses $t_n$ becomes a monotonically increasing function of $a$. However, even the classical radiation reaction force with quantum corrections can give an error in calculation of $t_n$ of about $40\%$. The greater $\chi$, the more probable the emission of a large part of particle energy is and the clearer  the stochasticity emerges (compare the solid and dash-dotted curves in Fig.\ref{fig:tn}).  The difference between $t_n$ obtained from equation (\ref{Ylong}) and from numerical calculation with allowance for radiation reaction (\ref{FrrQ}) is caused by the fact that Eq. (\ref{Ylong}) doesn't correctly describe motion close to the electric field node. However, far from the electric field node, the center of mass is described by $e^{\lambda_{\|+}t}$ when $a\gg a_{RR}$, even in the quasicalssical case, enabling calculation of  $\lambda_{\|+}$ as a function of $a$. In this case, for the considered wave amplitudes $\lambda_{\|+}$ is
\begin{equation}
\lambda_{\|+}^\mathrm{Q}\approx4a^{-0.3}.
\label{LlappQ}
\end{equation}
(see Fig.(\ref{fig:Rates})(a)). Comparison of $\lambda_{\|+}^\mathrm{Q}$ in the quasiclassical case and in the case with quantum corrected force shows that stochasticity doesn't change the power law but decreases the factor, $\lambda_{\|+}^\mathrm{Q}\approx0.73\lambda_{\|+}^\mathrm{corr}$.

Reduction of the rate of longitudinal drifting has a great impact on the development of QED cascades in the field of a circularly polarized standing wave. First, electrons (positrons) spend more time in the vicinity of electric field antinode and radiate more photons. Second, the longitudinal momentum of photons as well as of electrons is smaller due to radiation reaction, so displacement of the born pairs from the electric field antinode is smaller too. 

\section{Spatial distribution of $\mathrm{e^-e^+}$ plasma}

In this section we will pay particular attention to the spatial structures of pair plasmas created in vacuum through QED cascades in a standing circularly polarized wave, especially along the longitudinal direction. As QED cascades are generated in the high-field region, we can expect formation in colliding laser pulses of a hump-like density structure in the vicinity of these regions, at least in the avalanche regime of cascade development. This is not so, in general, because of a very important role of longitudinal particle drifting. However, in limiting cases of linear and circular polarization, the reasons are qualitatively different. NRT and at higher amplitudes ART regime can be realized in a linearly polarized standing wave, whereas in a standing wave with circular polarization only NRT regime occurs, drifting particles to the node region. We will give answers why and when different regimes of pair development occur, providing arguments that density distributions may be peaked at electric field antinode or node or in both regions.

At the stage of exponential growth of pairs it is natural to assume that hump-like density structures result in the competition of the growth rate of pair production and the corresponding particle escape rate from  the high-field region. Moreover, in a standing circularly polarized wave escaping particles are collected in the node region, as is seen in Fig.~\ref{mapQ}.    

Thus, to make estimates we have to compare three parameters. The first  parameter is the growth rate of electromagnetic cascade, which is maximal at the antinode where the field structure is a rotating electric field. Development of the cascade in such a field structure has been considered in detail in \cite{Elkina_PRSTAB2011,Nerush_POP2011,Fedotov_PRL2010}. The other two parameters are the rate of longitudinal particle drifting to the electric field node considered above, for which Eq. (\ref{LlappQ}) will be used as a more correct one, and the rate of transverse drifting.\par
\begin{figure}[h]
	\centering
		\includegraphics[width = 0.25\textwidth]{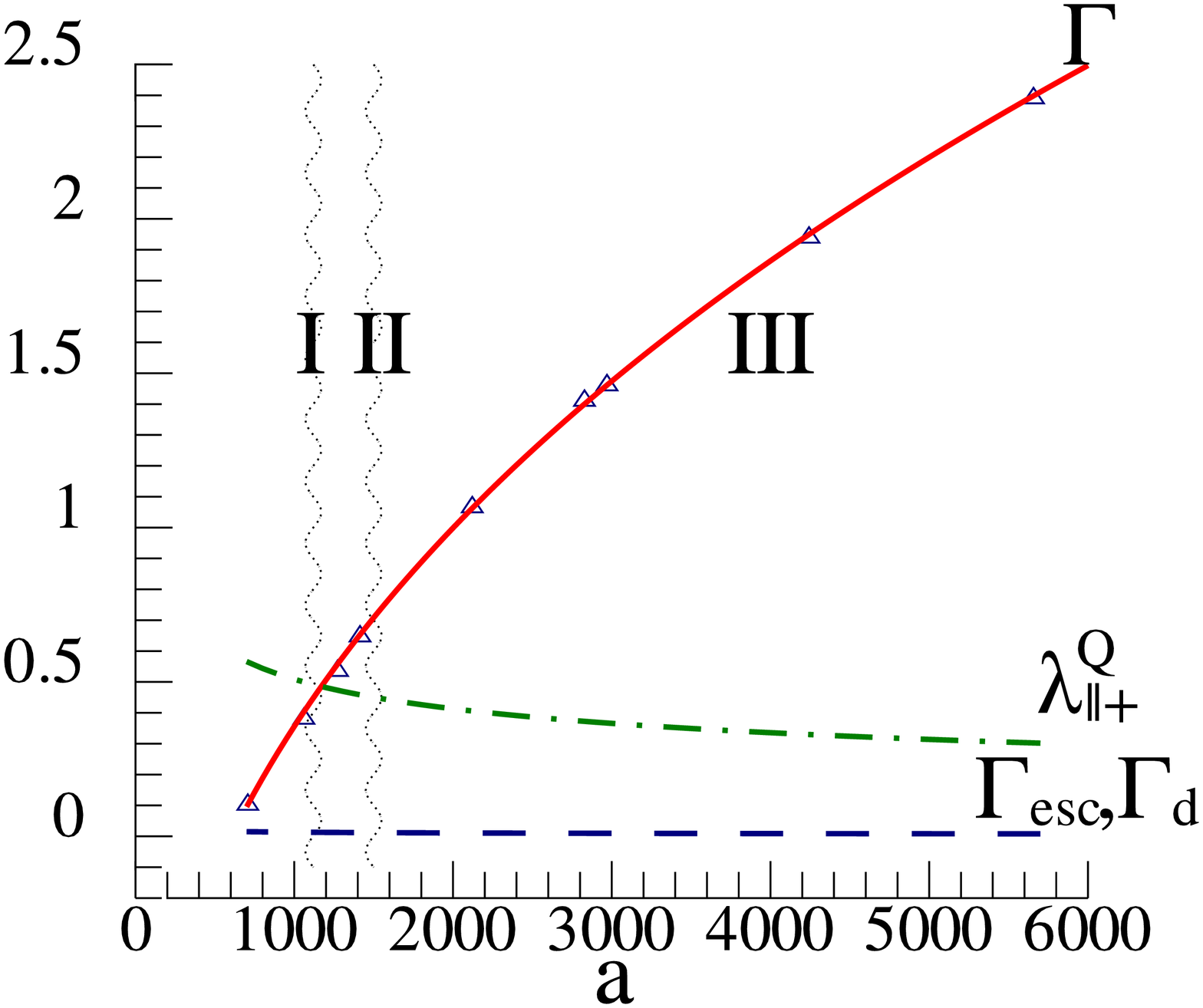}
	\caption{ (Color online) Electromagnetic cascade growth rate $\Gamma$ (solid line is approximation, triangles are obtained from numerical simulation), rate of longitudinal drift to electric field node $\lambda_{\|+}^\mathrm{Q}$ (dash-dotted line), rate of transverse drift $\Gamma_\mathrm{esc}$ and transverse diffusion drift $\Gamma_d$ at the electric field antinode (dashed line, $\mu\approx0.1$) versus field amplitude of two colliding circularly polarized laser beams. The dotted wavy lines separate regions of different pair plasma structures shown by Roman numerals.}
	\label{GrRt}
\end{figure}

Using the PIC-code PICADOR \cite{Bastrakov_JCS2012} that takes into account quantum effects in the frame of the quasiclassical approach \cite{Gonoskov_PRE2015} we calculated the cascade growth rate $\Gamma$ as a function of $a$ in the vicinity of the electric field antinode. The calculated $\Gamma$ is accurate to 0.22  with analytical approximation of cascade growth rate in the rotating electric field \cite{Nerush_POP2011}:
\begin{equation}
\Gamma=1.33\left(\frac{a}{a_{RR}}\right)^{1/4}\left(\frac{\tilde{\omega}}{\omega_l}\right)^{0.5}\lg\left(\frac{a}{a_{RR}}\right)-0.22,
\label{Gamma}
\end{equation}
where $\tilde{\omega}=2\pi c/10^{-4}=1.88\cdot10^{15}$~s$^{-1}$ (1$\mu$m wavelength).
The threshold amplitude for the cascade development is $a_\mathrm{cs}\approx650$, at this amplitude $\chi_{\mathrm{cs}}\approx0.87$. 

The transverse drifting implies two effects. Firstly, particles drift in inhomogeneous laser beams, as was considered in \cite{Fedotov_PRA2014}. Following this paper, in the case of weakly inhomogeneous field $E\propto a\exp(-\frac{\mu^2 r^2}{2})$ at $a\gg a_{RR}$ and $\mu\ll1$  close to the beam axis, the particle escape rate, assuming $r$ (that is of order beam radius $r_\mathrm{b}=\sqrt{\ln{2}}/\mu$ or less) as a function of time $r\approx r_0\exp\left(\Gamma_\mathrm{esc} t\right)$ is
\begin{equation}
\Gamma_\mathrm{esc}\approx\frac{5}{8}\mu^2\left(\frac{a_{RR}}{a}\right)^{0.75}.
\label{GEsc}
\end{equation}
The field amplitude of one beam is $a/2$.
This conclusion is valid for large beams with radius $r_\mathrm{b}\gtrsim3\lambda$, which corresponds to $\mu\lesssim0.1$. For these parameters $\Gamma_\mathrm{esc}$ is much less than $\Gamma$ and $\lambda_{\|+}^\mathrm{LL}$ as well, and transverse drifting can be considered independently and almost doesn't change the longitudinal drifting. Although this conclusion is analytically proved in the frame of LL(LAD) force, it is also valid for the quantum-corrected force and in the quasiclassical case. 

In the case of LL force $\Gamma_\mathrm{esc}/\lambda_{\|+}^\mathrm{LL}=\frac{5\ln(2)}{8r_b^2}=\frac{U}{r_b^2}$ as follows from Eqs.(\ref{FrrG}),(\ref{Llapp}), where $U=0.43\sim1$. In dimensional variables $\Gamma_\mathrm{esc}/\lambda_{\|+}^\mathrm{LL}=U/(k_lr_b)^2$ ($k_l=\omega_l/c$) and doesn't depend on $a$. These two rates  are specified by ponderomotive force, but in different directions. The characteristic time of escape is proportional to the ratio of the inhomogeneity scale to drift velocity. Moreover, the velocity is proportional to the field gradient, thus inversely proportional to the inhomogeneity scale. The scale is $1/k_l$ in the longitudinal direction, and  $r_b$ in the transverse direction. Thereby in any case (LL, LAD, quantum corrected forces, quasiclassical case) we can state that $\Gamma_\mathrm{esc}/\lambda_{\|+}=U/(k_lr_b)^2$. So, even in a tightly focused field $\Gamma_\mathrm{esc}<\lambda_{\|+}$. Numerical simulations confirm this conclusion. 

Another drifting effect is diffusion due to the stochastic nature of photon emission considered in Section \ref{Sec3}. Its rate is $\Gamma_d=4d/r_b^2$. When $\chi\gg1$ from Eqs.(\ref{GLLQ}),(\ref{Diff}) it follows that
\begin{equation}
\Gamma_d\approx\frac{2\eta^{1/3}}{3\alpha r_b^2}\left(\frac{\delta_q}{a}\right)^{(1-s)/3}.
\label{Gd}
\end{equation}
In the quantum case $s=0.25$, $\delta_q\approx0.37\alpha/\eta^{1/3}$ this gives asymptotically $\Gamma_d/\lambda_{\|+}^\mathrm{Q}=0.24/r_b^2$. Diffusion drifting is the order of magnitude of drifting due to field inhomogeneity $\Gamma_d\sim\Gamma_\mathrm{esc}$ (in the quantum case $\chi\gg1$) but they are also less important than longitudinal drifting, even for tightly focused laser beams.

In Fig.~\ref{GrRt} we summarize all parameters needed for QED cascade development analysis as a function of field amplitude. First of all, we determine the point where
\begin{equation} 
\Gamma(a)=\lambda_{\|+}^\mathrm{Q}(a),
\label{EqTh1}
\end{equation} 
i.e. avalanche growth rate is exactly compensated by the particle escape rate. The solution of equation (\ref{EqTh1}), which is
\begin{equation}
a=a_{\mathrm{th}1}\approx1150,
\label{ath1}
\end{equation}
defines the threshold of cascade development for the continuous wave, i.e. for $\mathrm{e^-e^+}$ plasma production it should be $a>a_{\mathrm{th}1}$. Next, we consider pair production in two regions: antinode and node of electric field. In the antinode region the pair production rate is $\Gamma(a)-\lambda_{\|+}^\mathrm{Q}(a)$ and the plasma density as a function of time is $n_a(t)=n_0e^{(\Gamma-\lambda_{\|+}^\mathrm{Q})t}$. $n_0$ is initial plasma density. In the node region the particle growth is $\lambda_{\|+}^\mathrm{Q}(a)$, as the particles escaping from the antinode drift exactly to the node as shown in Fig.~\ref{mapQ}. However, the particles need time $\delta t\approx{\lambda_{\|+}^\mathrm{Q}}^{-1}$ to reach the node region, so $dn_n/dt=\lambda_{\|+}^\mathrm{Q}n_a(t-\delta t)$, and plasma density in the node region is
\begin{equation}
n_n(t)=\frac{\lambda_{\|+}^\mathrm{Q}}{\Gamma-\lambda_{\|+}^\mathrm{Q}}n_0e^{(\Gamma-\lambda_{\|+}^\mathrm{Q})(t-\delta t)}.
\label{nnt}
\end{equation}
Density peaks in the node and antinode regions are the same, when $n_a(t)\approx n_n(t)$. Consequently 
\begin{equation}
(\Gamma/\lambda_{\|+}^\mathrm{Q}-1)e^{\Gamma/\lambda_{\|+}^\mathrm{Q}-1}\approx1,
\label{EqTh2}
\end{equation}
having a solution
\begin{equation}
\Gamma\approx1.57\lambda_{\|+}^\mathrm{Q}.
\label{EqTh22}
\end{equation}
As follows from Fig.~\ref{GrRt} Eq.~(\ref{EqTh22}) is satisfied at
\begin{equation} 
a=a_{\mathrm{th}2}\approx1500.
\label{ath2}
\end{equation}
 Thus we can identify three regimes of electromagnetic cascade development. In the case when the intensities just slightly exceed the threshold $0<\Gamma(a)-\lambda_{\|+}^\mathrm{Q}(a)\ll\lambda_{\|+}^\mathrm{Q}(a)$ pairs are located mainly in the vicinity of the node. We mark this regime as first regime I in Fig.~\ref{GrRt}. At the intensities near the second threshold, pairs are located in both antinode and node regions with comparable peak density values (second regime II). And the third regime (III) occurs at higher intensities $a>a_{\mathrm{th}2}$ when the peaks in the node region have a lower density than in the antinode region. In this case, at much higher intensities when $\Gamma(a)\gg\lambda_{\|+}^\mathrm{Q}(a)$, density distribution will peak around the antinode plane only.   

By using the PIC-code PICADOR we performed 3D simulations of cascade development in counter-propagating circularly polarized laser beams, with a focus on the avalanche regime when plasma back reaction is negligible. Laser pulses are half infinite with one wave period leading edge. The initial plasma seed density was very low $n_0=0.01$cm$^{-3}$ to omit plasma back reaction. To eliminate the influence of the leading edge, electrons and positrons appear in numerical simulation in the field region $0.1\lambda_l\times0.01\lambda_l\times0.1\lambda_l$ around the point of the maximum of electric field amplitude when a standing wave is formed. This is reasonable because two counter-propagating circularly polarized laser pulses can strongly compress plasma target. The simulation box included $224\times128\times224$ cells and was 3D  as $7\lambda_l\times2\lambda_l\times7\lambda_l$, the time step was $1/(32\omega_l)$. We performed a parametric scan for a wide range of incident amplitudes. Analysis of the cascade development revealed three different regimes of pair plasma evolution resulting in three types of spatial density structures.
\begin{figure}[h]
	\centering
		\includegraphics[width = 0.155\textwidth]{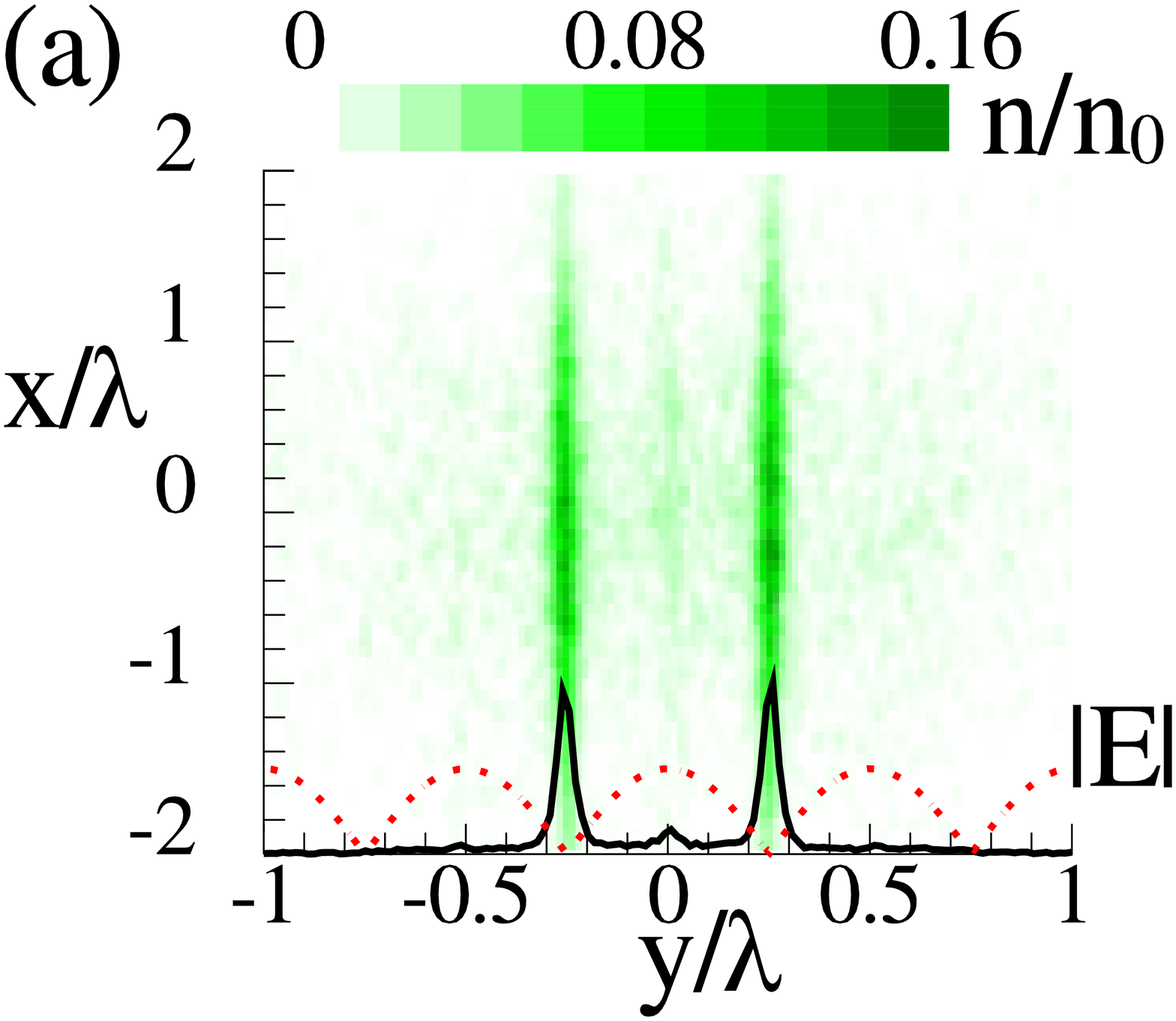}
		\includegraphics[width = 0.155\textwidth]{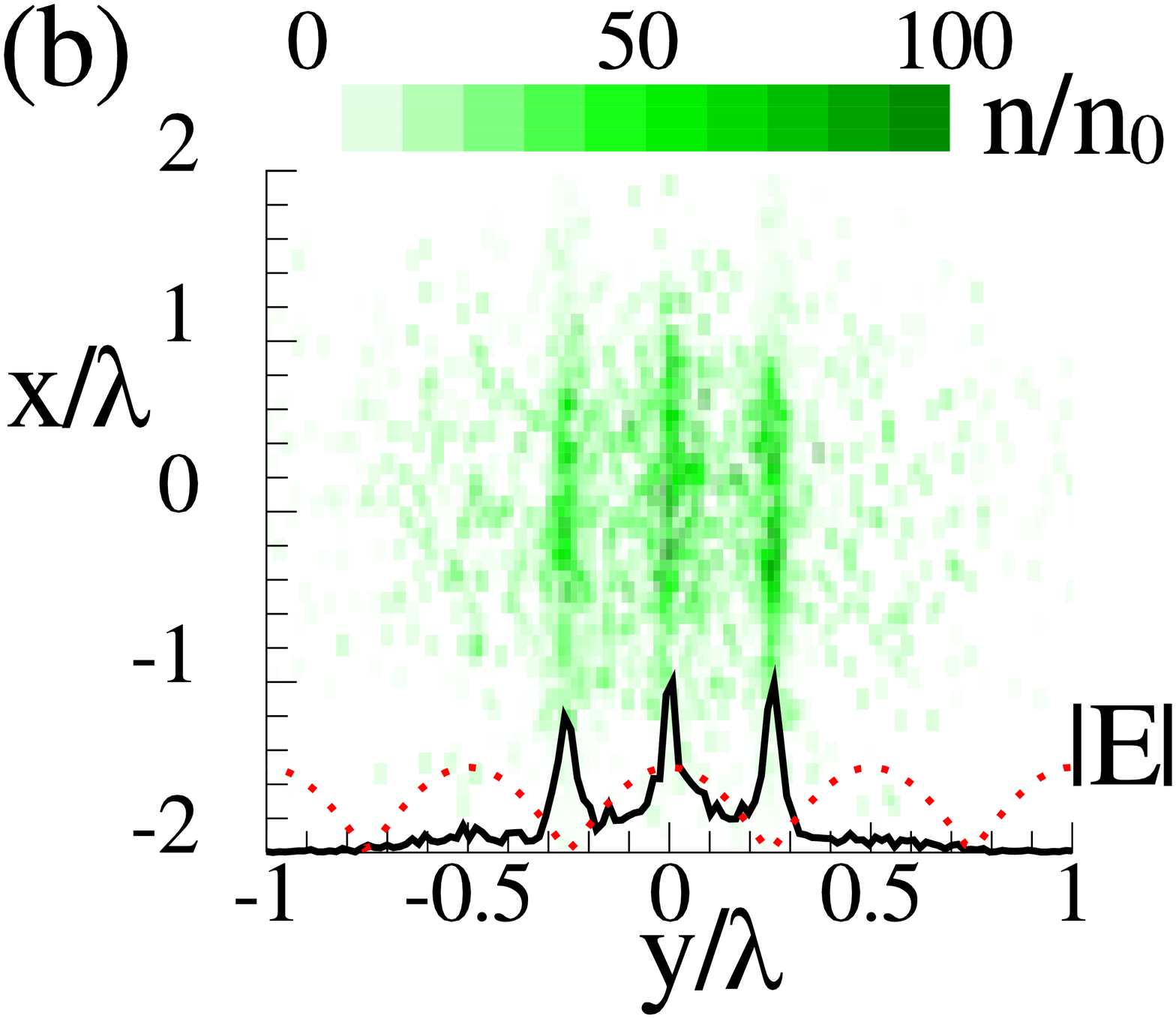}
		\includegraphics[width = 0.155\textwidth]{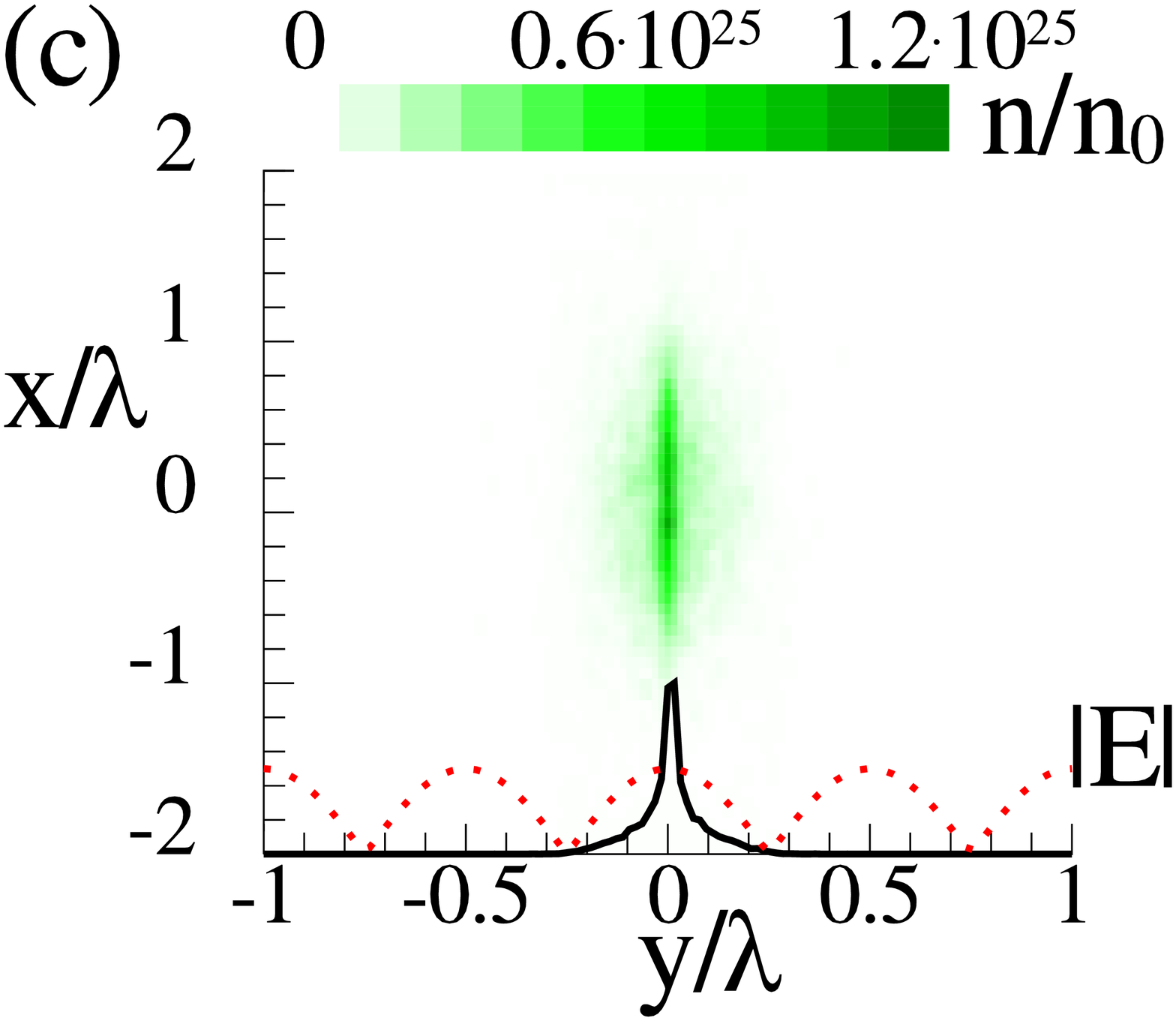}
	\caption{(Color online) Pair plasma structures at the initial stage of electromagnetic cascades in the field of two colliding laser beams with $3\lambda_l$ diameter at FWHM and amplitudes (a) $a/2=550$, (b) $a/2=700$, (c) $a/2=2500$. Electron density $n$ is normalized to initial electron density $n_0$. Dotted line represents electric field magnitude along symmetry axis of beams $r=0$. Profile of electron distribution averaged along $x$ direction is shown by solid black line.}
	\label{CasStruc}
\end{figure}
As follows from the simulations, the first threshold, when cascades start to develop, is about $a_{\mathrm{th}1}\approx1100$ and the second one, which we defined at the wave amplitude when maximum values of the total pair number over transverse beam section (black solid curves in Fig.\ref{CasStruc}) are approximately equal, is $a_{\mathrm{th}2}\approx1400$. These threshold values are quite close to those obtained above. According to the numerical simulations, the peaks in the node region disappear when $a>1900$ and $\Gamma>2.25\lambda_{\|+}^\mathrm{Q}$. In Fig.\ref{CasStruc} we present typical results of 3D simulations for the laser beam radius  $r_b=1.5\lambda_l$ ($\mu\approx0.1$) in the form of pair plasma distribution after 5 laser periods for amplitudes $a/2=550\mathrm{(a)};700\mathrm{(b)};2500\mathrm{(c)}$. It should be mentioned that there is also transverse drift in the node region. Unlike the antinode region radiation losses don't suppress transverse drift in the node region, because particle energy is small $\gamma\ll a$ and $F_r\ll1$. The particles move in the NRT regime there and, as was considered in Sec.\ref{Sec2}, drift transversely with velocity of about $0.6c$. The corresponding characteristic rate of escape $\Gamma_{d}^n$ in dimensionless variables is $0.6/r_b$. For the considered parameters $\Gamma_{d}^n=0.064$ is much less than $\lambda_{\|+}^\mathrm{Q}$ and this drift can be neglected. Clearly, the results of simulations are consistent with the three regimes discussed above.

\section{Summary}

In this paper we tried to understand what types of spatial $\mathrm{e^-e^+}$ plasma structures may be realized through QED cascades in counter-propagating laser pulses with circular polarization. The qualitative difference from the case of linearly polarized pulses makes this problem interesting and fundamentally important for understanding QED plasma dynamics in laser fields. To get an insight into the physics, we first presented long-term density distributions, in which asymptotic regimes such as ponderomotive trapping and relativistic chaos are inherently included. It was shown that only the NRT regime is realized taking into account the radiation reaction effect, whereas the ART regime, trapping electrons in the vicinity of electric field antinode, is crucially important for linear polarization. Since QED cascades are mainly generated in the high-field region we presented a general analysis of longitudinal particle drifting at extreme intensities when the quantum radiation reaction effect should be accounted for. For qualitative estimation we also considered the stochastic nature of photon emission, particularly showing that discreteness of emission can additionally decrease drifting rates up to 1.4 times due to strong perturbation of particle motion  and generate a new effect of particle diffusion. Based on the comparison of pair production growth rates and the main particle loss rates connected with longitudinal drifting from electric field antinode to the node we conclude that three modes of QED cascades may be formed in a standing circularly polarized wave, giving rise to density distributions peaked at antinode or node or in both regions. This conclusion is confirmed by PIC simulations.     

The authors acknowledge support from the Russian Science Foundation project No. 16-12-10486 (analytical part of the work), the Russian Foundation for Basic Research project No. 15-32-20641 (numerical simulations). A.V.B. acknowledges the Dynasty Foundation support.
\section{Appendix}
\label{AppFr}
For simplicity of numerical calculations it is possible to use the following expression for quantum-corrected RR force which corresponds to (\ref{FrrQ}) to an accuracy within $0.15\%$ 
\begin{equation}
\label{FrApp1}
F_{ra}\approx\frac{\alpha}{3\sqrt{3}\pi\gamma\eta}
\begin{cases}
\begin{aligned}
10.8828\chi^2&\left(1+18.08\chi+68.7\chi^2+70.8\chi^3+\right.\\
&\left.7.6403\chi^4\right)^{-1/3}\end{aligned} & \text{if $\chi<10$;}\\
\begin{aligned}
-10.8828+&6.05498\chi^{2/3}+28.551\chi^{-2/3}-\\
&41.469\chi^{-1}+24.7245\chi^{-4/3}-8.1621\chi^{-2}\end{aligned} & \text{if $\chi\ge10$.}

\end{cases}
\end{equation}
\bibliography{Bibl1}

\begin{thebibliography}{38}%
\makeatletter
\providecommand \@ifxundefined [1]{%
 \@ifx{#1\undefined}
}%
\providecommand \@ifnum [1]{%
 \ifnum #1\expandafter \@firstoftwo
 \else \expandafter \@secondoftwo
 \fi
}%
\providecommand \@ifx [1]{%
 \ifx #1\expandafter \@firstoftwo
 \else \expandafter \@secondoftwo
 \fi
}%
\providecommand \natexlab [1]{#1}%
\providecommand \enquote  [1]{``#1''}%
\providecommand \bibnamefont  [1]{#1}%
\providecommand \bibfnamefont [1]{#1}%
\providecommand \citenamefont [1]{#1}%
\providecommand \href@noop [0]{\@secondoftwo}%
\providecommand \href [0]{\begingroup \@sanitize@url \@href}%
\providecommand \@href[1]{\@@startlink{#1}\@@href}%
\providecommand \@@href[1]{\endgroup#1\@@endlink}%
\providecommand \@sanitize@url [0]{\catcode `\\12\catcode `\$12\catcode
  `\&12\catcode `\#12\catcode `\^12\catcode `\_12\catcode `\%12\relax}%
\providecommand \@@startlink[1]{}%
\providecommand \@@endlink[0]{}%
\providecommand \url  [0]{\begingroup\@sanitize@url \@url }%
\providecommand \@url [1]{\endgroup\@href {#1}{\urlprefix }}%
\providecommand \urlprefix  [0]{URL }%
\providecommand \Eprint [0]{\href }%
\providecommand \doibase [0]{http://dx.doi.org/}%
\providecommand \selectlanguage [0]{\@gobble}%
\providecommand \bibinfo  [0]{\@secondoftwo}%
\providecommand \bibfield  [0]{\@secondoftwo}%
\providecommand \translation [1]{[#1]}%
\providecommand \BibitemOpen [0]{}%
\providecommand \bibitemStop [0]{}%
\providecommand \bibitemNoStop [0]{.\EOS\space}%
\providecommand \EOS [0]{\spacefactor3000\relax}%
\providecommand \BibitemShut  [1]{\csname bibitem#1\endcsname}%
\let\auto@bib@innerbib\@empty
\bibitem [{ELI()}]{ELI}%
  \BibitemOpen
  \href@noop {} {}\bibinfo {howpublished}
  {http://www.eli-beams.eu/}\BibitemShut {NoStop}%
\bibitem [{\citenamefont {Zou}\ \emph {et~al.}(2015)\citenamefont {Zou},
  \citenamefont {Le~Blanc}, \citenamefont {Papadopoulos}, \citenamefont
  {Cheriaux}, \citenamefont {Georges}, \citenamefont {Mennerat}, \citenamefont
  {Druon}, \citenamefont {Lecherbourg}, \citenamefont {Pellegrina},
  \citenamefont {Ramirez}, \citenamefont {Giambruno}, \citenamefont {Freneaux},
  \citenamefont {Leconte}, \citenamefont {Badarau}, \citenamefont {Boudenne},
  \citenamefont {Fournet}, \citenamefont {Valloton}, \citenamefont {Paillard},
  \citenamefont {Veray}, \citenamefont {Pina}, \citenamefont {Monot},
  \citenamefont {Chambaret}, \citenamefont {Martin}, \citenamefont {Mathieu},
  \citenamefont {Audebert},\ and\ \citenamefont {Amiranoff}}]{Appolon10}%
  \BibitemOpen
  \bibfield  {author} {\bibinfo {author} {\bibfnamefont {J.}~\bibnamefont
  {Zou}}, \bibinfo {author} {\bibfnamefont {C.}~\bibnamefont {Le~Blanc}},
  \bibinfo {author} {\bibfnamefont {D.}~\bibnamefont {Papadopoulos}}, \bibinfo
  {author} {\bibfnamefont {G.}~\bibnamefont {Cheriaux}}, \bibinfo {author}
  {\bibfnamefont {P.}~\bibnamefont {Georges}}, \bibinfo {author} {\bibfnamefont
  {G.}~\bibnamefont {Mennerat}}, \bibinfo {author} {\bibfnamefont
  {F.}~\bibnamefont {Druon}}, \bibinfo {author} {\bibfnamefont
  {L.}~\bibnamefont {Lecherbourg}}, \bibinfo {author} {\bibfnamefont
  {A.}~\bibnamefont {Pellegrina}}, \bibinfo {author} {\bibfnamefont
  {P.}~\bibnamefont {Ramirez}}, \bibinfo {author} {\bibfnamefont
  {F.}~\bibnamefont {Giambruno}}, \bibinfo {author} {\bibfnamefont
  {A.}~\bibnamefont {Freneaux}}, \bibinfo {author} {\bibfnamefont
  {F.}~\bibnamefont {Leconte}}, \bibinfo {author} {\bibfnamefont
  {D.}~\bibnamefont {Badarau}}, \bibinfo {author} {\bibfnamefont
  {J.}~\bibnamefont {Boudenne}}, \bibinfo {author} {\bibfnamefont
  {D.}~\bibnamefont {Fournet}}, \bibinfo {author} {\bibfnamefont
  {T.}~\bibnamefont {Valloton}}, \bibinfo {author} {\bibfnamefont
  {J.}~\bibnamefont {Paillard}}, \bibinfo {author} {\bibfnamefont
  {J.}~\bibnamefont {Veray}}, \bibinfo {author} {\bibfnamefont
  {M.}~\bibnamefont {Pina}}, \bibinfo {author} {\bibfnamefont {P.}~\bibnamefont
  {Monot}}, \bibinfo {author} {\bibfnamefont {J.}~\bibnamefont {Chambaret}},
  \bibinfo {author} {\bibfnamefont {P.}~\bibnamefont {Martin}}, \bibinfo
  {author} {\bibfnamefont {F.}~\bibnamefont {Mathieu}}, \bibinfo {author}
  {\bibfnamefont {P.}~\bibnamefont {Audebert}}, \ and\ \bibinfo {author}
  {\bibfnamefont {F.}~\bibnamefont {Amiranoff}},\ }\href@noop {} {\bibfield
  {journal} {\bibinfo  {journal} {High Power Laser Science and Engineering}\
  }\textbf {\bibinfo {volume} {3}},\ \bibinfo {pages} {e2 (4 pages)} (\bibinfo
  {year} {2015})}\BibitemShut {NoStop}%
\bibitem [{\citenamefont {Bashinov}\ \emph {et~al.}(2014)\citenamefont
  {Bashinov}, \citenamefont {Gonoskov}, \citenamefont {Kim}, \citenamefont
  {Mourou},\ and\ \citenamefont {Sergeev}}]{XCELSa}%
  \BibitemOpen
  \bibfield  {author} {\bibinfo {author} {\bibfnamefont {A.~V.}\ \bibnamefont
  {Bashinov}}, \bibinfo {author} {\bibfnamefont {A.~A.}\ \bibnamefont
  {Gonoskov}}, \bibinfo {author} {\bibfnamefont {A.~V.}\ \bibnamefont {Kim}},
  \bibinfo {author} {\bibfnamefont {G.}~\bibnamefont {Mourou}}, \ and\ \bibinfo
  {author} {\bibfnamefont {A.~M.}\ \bibnamefont {Sergeev}},\ }\href {\doibase
  10.1140/epjst/e2014-02161-7} {\bibfield  {journal} {\bibinfo  {journal} {Eur.
  Phys. J. Special Topics}\ }\textbf {\bibinfo {volume} {223}},\ \bibinfo
  {pages} {1105} (\bibinfo {year} {2014})}\BibitemShut {NoStop}%
\bibitem [{\citenamefont {Tamburini}\ \emph {et~al.}(2011)\citenamefont
  {Tamburini}, \citenamefont {Pegoraro}, \citenamefont {Piazza}, \citenamefont
  {Keitel}, \citenamefont {Liseykina},\ and\ \citenamefont
  {Macchi}}]{Tamburini_NuclInst}%
  \BibitemOpen
  \bibfield  {author} {\bibinfo {author} {\bibfnamefont {M.}~\bibnamefont
  {Tamburini}}, \bibinfo {author} {\bibfnamefont {F.}~\bibnamefont {Pegoraro}},
  \bibinfo {author} {\bibfnamefont {A.~D.}\ \bibnamefont {Piazza}}, \bibinfo
  {author} {\bibfnamefont {C.}~\bibnamefont {Keitel}}, \bibinfo {author}
  {\bibfnamefont {T.}~\bibnamefont {Liseykina}}, \ and\ \bibinfo {author}
  {\bibfnamefont {A.}~\bibnamefont {Macchi}},\ }\href {\doibase
  http://dx.doi.org/10.1016/j.nima.2010.12.056} {\bibfield  {journal} {\bibinfo
   {journal} {Nucl. Instrum. Methods Phys. Res., Sec. A}\ }\textbf {\bibinfo
  {volume} {653}},\ \bibinfo {pages} {181 } (\bibinfo {year}
  {2011})}\BibitemShut {NoStop}%
\bibitem [{\citenamefont {Gonoskov}\ \emph {et~al.}(2014)\citenamefont
  {Gonoskov}, \citenamefont {Bashinov}, \citenamefont {Gonoskov}, \citenamefont
  {Harvey}, \citenamefont {Ilderton}, \citenamefont {Kim}, \citenamefont
  {Marklund}, \citenamefont {Mourou},\ and\ \citenamefont
  {Sergeev}}]{ART_Gonoskov}%
  \BibitemOpen
  \bibfield  {author} {\bibinfo {author} {\bibfnamefont {A.}~\bibnamefont
  {Gonoskov}}, \bibinfo {author} {\bibfnamefont {A.}~\bibnamefont {Bashinov}},
  \bibinfo {author} {\bibfnamefont {I.}~\bibnamefont {Gonoskov}}, \bibinfo
  {author} {\bibfnamefont {C.}~\bibnamefont {Harvey}}, \bibinfo {author}
  {\bibfnamefont {A.}~\bibnamefont {Ilderton}}, \bibinfo {author}
  {\bibfnamefont {A.}~\bibnamefont {Kim}}, \bibinfo {author} {\bibfnamefont
  {M.}~\bibnamefont {Marklund}}, \bibinfo {author} {\bibfnamefont
  {G.}~\bibnamefont {Mourou}}, \ and\ \bibinfo {author} {\bibfnamefont
  {A.}~\bibnamefont {Sergeev}},\ }\href {\doibase
  10.1103/PhysRevLett.113.014801} {\bibfield  {journal} {\bibinfo  {journal}
  {Phys. Rev. Lett.}\ }\textbf {\bibinfo {volume} {113}},\ \bibinfo {pages}
  {014801} (\bibinfo {year} {2014})}\BibitemShut {NoStop}%
\bibitem [{\citenamefont {Ji}\ \emph {et~al.}(2014)\citenamefont {Ji},
  \citenamefont {Pukhov}, \citenamefont {Kostyukov}, \citenamefont {Shen},\
  and\ \citenamefont {Akli}}]{RT_Pukhov}%
  \BibitemOpen
  \bibfield  {author} {\bibinfo {author} {\bibfnamefont {L.~L.}\ \bibnamefont
  {Ji}}, \bibinfo {author} {\bibfnamefont {A.}~\bibnamefont {Pukhov}}, \bibinfo
  {author} {\bibfnamefont {I.~Y.}\ \bibnamefont {Kostyukov}}, \bibinfo {author}
  {\bibfnamefont {B.~F.}\ \bibnamefont {Shen}}, \ and\ \bibinfo {author}
  {\bibfnamefont {K.}~\bibnamefont {Akli}},\ }\href {\doibase
  10.1103/PhysRevLett.112.145003} {\bibfield  {journal} {\bibinfo  {journal}
  {Phys. Rev. Lett.}\ }\textbf {\bibinfo {volume} {112}},\ \bibinfo {pages}
  {145003} (\bibinfo {year} {2014})}\BibitemShut {NoStop}%
\bibitem [{\citenamefont {Duclous}\ \emph {et~al.}(2011)\citenamefont
  {Duclous}, \citenamefont {Kirk},\ and\ \citenamefont
  {Bell}}]{Duclous_PPCF2011}%
  \BibitemOpen
  \bibfield  {author} {\bibinfo {author} {\bibfnamefont {R.}~\bibnamefont
  {Duclous}}, \bibinfo {author} {\bibfnamefont {J.~G.}\ \bibnamefont {Kirk}}, \
  and\ \bibinfo {author} {\bibfnamefont {A.~R.}\ \bibnamefont {Bell}},\ }\href
  {http://stacks.iop.org/0741-3335/53/i=1/a=015009} {\bibfield  {journal}
  {\bibinfo  {journal} {Plasma Physics and Controlled Fusion}\ }\textbf
  {\bibinfo {volume} {53}},\ \bibinfo {pages} {015009} (\bibinfo {year}
  {2011})}\BibitemShut {NoStop}%
\bibitem [{\citenamefont {Neitz}\ and\ \citenamefont
  {Di~Piazza}(2013)}]{Neitz_PRL}%
  \BibitemOpen
  \bibfield  {author} {\bibinfo {author} {\bibfnamefont {N.}~\bibnamefont
  {Neitz}}\ and\ \bibinfo {author} {\bibfnamefont {A.}~\bibnamefont
  {Di~Piazza}},\ }\href {\doibase 10.1103/PhysRevLett.111.054802} {\bibfield
  {journal} {\bibinfo  {journal} {Phys. Rev. Lett.}\ }\textbf {\bibinfo
  {volume} {111}},\ \bibinfo {pages} {054802} (\bibinfo {year}
  {2013})}\BibitemShut {NoStop}%
\bibitem [{\citenamefont {Yoffe}\ \emph {et~al.}(2015)\citenamefont {Yoffe},
  \citenamefont {Kravets}, \citenamefont {Noble},\ and\ \citenamefont
  {Jaroszynski}}]{Yoffe_NJP2015}%
  \BibitemOpen
  \bibfield  {author} {\bibinfo {author} {\bibfnamefont {S.}~\bibnamefont
  {Yoffe}}, \bibinfo {author} {\bibfnamefont {Y.}~\bibnamefont {Kravets}},
  \bibinfo {author} {\bibfnamefont {A.}~\bibnamefont {Noble}}, \ and\ \bibinfo
  {author} {\bibfnamefont {D.~A.}\ \bibnamefont {Jaroszynski}},\ }\href
  {http://stacks.iop.org/1367-2630/17/i=5/a=053025} {\bibfield  {journal}
  {\bibinfo  {journal} {New J. Phys.}\ }\textbf {\bibinfo {volume} {17}},\
  \bibinfo {pages} {053025} (\bibinfo {year} {2015})}\BibitemShut {NoStop}%
\bibitem [{\citenamefont {Bashinov}\ \emph {et~al.}(2015)\citenamefont
  {Bashinov}, \citenamefont {Kim},\ and\ \citenamefont
  {Sergeev}}]{Bashinov_PRE2015}%
  \BibitemOpen
  \bibfield  {author} {\bibinfo {author} {\bibfnamefont {A.~V.}\ \bibnamefont
  {Bashinov}}, \bibinfo {author} {\bibfnamefont {A.~V.}\ \bibnamefont {Kim}}, \
  and\ \bibinfo {author} {\bibfnamefont {A.~M.}\ \bibnamefont {Sergeev}},\
  }\href {\doibase 10.1103/PhysRevE.92.043105} {\bibfield  {journal} {\bibinfo
  {journal} {Phys. Rev. E}\ }\textbf {\bibinfo {volume} {92}},\ \bibinfo
  {pages} {043105} (\bibinfo {year} {2015})}\BibitemShut {NoStop}%
\bibitem [{\citenamefont {Breit}\ and\ \citenamefont
  {Wheeler}(1934)}]{Breit_PR1934}%
  \BibitemOpen
  \bibfield  {author} {\bibinfo {author} {\bibfnamefont {G.}~\bibnamefont
  {Breit}}\ and\ \bibinfo {author} {\bibfnamefont {J.~A.}\ \bibnamefont
  {Wheeler}},\ }\href {\doibase 10.1103/PhysRev.46.1087} {\bibfield  {journal}
  {\bibinfo  {journal} {Phys. Rev.}\ }\textbf {\bibinfo {volume} {46}},\
  \bibinfo {pages} {1087} (\bibinfo {year} {1934})}\BibitemShut {NoStop}%
\bibitem [{\citenamefont {Narozhny}\ \emph {et~al.}(1965)\citenamefont
  {Narozhny}, \citenamefont {Nikishov},\ and\ \citenamefont
  {Ritus}}]{Narozhny_JETP1965}%
  \BibitemOpen
  \bibfield  {author} {\bibinfo {author} {\bibfnamefont {N.~B.}\ \bibnamefont
  {Narozhny}}, \bibinfo {author} {\bibfnamefont {A.~I.}\ \bibnamefont
  {Nikishov}}, \ and\ \bibinfo {author} {\bibfnamefont {V.~I.}\ \bibnamefont
  {Ritus}},\ }\href@noop {} {\bibfield  {journal} {\bibinfo  {journal} {Sov.
  Phys. JETP}\ }\textbf {\bibinfo {volume} {20}} (\bibinfo {year}
  {1965})}\BibitemShut {NoStop}%
\bibitem [{\citenamefont {Bell}\ and\ \citenamefont
  {Kirk}(2008)}]{Bell_PRL2008}%
  \BibitemOpen
  \bibfield  {author} {\bibinfo {author} {\bibfnamefont {A.~R.}\ \bibnamefont
  {Bell}}\ and\ \bibinfo {author} {\bibfnamefont {J.~G.}\ \bibnamefont
  {Kirk}},\ }\href {\doibase 10.1103/PhysRevLett.101.200403} {\bibfield
  {journal} {\bibinfo  {journal} {Phys. Rev. Lett.}\ }\textbf {\bibinfo
  {volume} {101}},\ \bibinfo {pages} {200403} (\bibinfo {year}
  {2008})}\BibitemShut {NoStop}%
\bibitem [{\citenamefont {Nerush}\ \emph
  {et~al.}(2011{\natexlab{a}})\citenamefont {Nerush}, \citenamefont
  {Kostyukov}, \citenamefont {Fedotov}, \citenamefont {Narozhny}, \citenamefont
  {Elkina},\ and\ \citenamefont {Ruhl}}]{Nerush_PRL2011}%
  \BibitemOpen
  \bibfield  {author} {\bibinfo {author} {\bibfnamefont {E.~N.}\ \bibnamefont
  {Nerush}}, \bibinfo {author} {\bibfnamefont {I.~Y.}\ \bibnamefont
  {Kostyukov}}, \bibinfo {author} {\bibfnamefont {A.~M.}\ \bibnamefont
  {Fedotov}}, \bibinfo {author} {\bibfnamefont {N.~B.}\ \bibnamefont
  {Narozhny}}, \bibinfo {author} {\bibfnamefont {N.~V.}\ \bibnamefont
  {Elkina}}, \ and\ \bibinfo {author} {\bibfnamefont {H.}~\bibnamefont
  {Ruhl}},\ }\href {\doibase 10.1103/PhysRevLett.106.035001} {\bibfield
  {journal} {\bibinfo  {journal} {Phys. Rev. Lett.}\ }\textbf {\bibinfo
  {volume} {106}},\ \bibinfo {pages} {035001} (\bibinfo {year}
  {2011}{\natexlab{a}})}\BibitemShut {NoStop}%
\bibitem [{\citenamefont {Bashmakov}\ \emph {et~al.}(2014)\citenamefont
  {Bashmakov}, \citenamefont {Nerush}, \citenamefont {Kostyukov}, \citenamefont
  {Fedotov},\ and\ \citenamefont {Narozhny}}]{Bashmakov_POP2014}%
  \BibitemOpen
  \bibfield  {author} {\bibinfo {author} {\bibfnamefont {V.~F.}\ \bibnamefont
  {Bashmakov}}, \bibinfo {author} {\bibfnamefont {E.~N.}\ \bibnamefont
  {Nerush}}, \bibinfo {author} {\bibfnamefont {I.~Y.}\ \bibnamefont
  {Kostyukov}}, \bibinfo {author} {\bibfnamefont {A.~M.}\ \bibnamefont
  {Fedotov}}, \ and\ \bibinfo {author} {\bibfnamefont {N.~B.}\ \bibnamefont
  {Narozhny}},\ }\href {\doibase 10.1063/1.4861863} {\bibfield  {journal}
  {\bibinfo  {journal} {Phys. Plasmas}\ }\textbf {\bibinfo {volume} {21}},\
  \bibinfo {pages} {013105} (\bibinfo {year} {2014})}\BibitemShut {NoStop}%
\bibitem [{\citenamefont {Grismayer}\ \emph {et~al.}(2016)\citenamefont
  {Grismayer}, \citenamefont {Vranic}, \citenamefont {Martins}, \citenamefont
  {Fonseca},\ and\ \citenamefont {Silva}}]{Grismayer_POP2016}%
  \BibitemOpen
  \bibfield  {author} {\bibinfo {author} {\bibfnamefont {T.}~\bibnamefont
  {Grismayer}}, \bibinfo {author} {\bibfnamefont {M.}~\bibnamefont {Vranic}},
  \bibinfo {author} {\bibfnamefont {J.~L.}\ \bibnamefont {Martins}}, \bibinfo
  {author} {\bibfnamefont {R.~A.}\ \bibnamefont {Fonseca}}, \ and\ \bibinfo
  {author} {\bibfnamefont {L.~O.}\ \bibnamefont {Silva}},\ }\href {\doibase
  10.1063/1.4950841} {\bibfield  {journal} {\bibinfo  {journal} {Phys.
  Plasmas}\ }\textbf {\bibinfo {volume} {23}},\ \bibinfo {pages} {056706}
  (\bibinfo {year} {2016})}\BibitemShut {NoStop}%
\bibitem [{\citenamefont {Jirka}\ \emph {et~al.}(2016)\citenamefont {Jirka},
  \citenamefont {Klimo}, \citenamefont {Bulanov}, \citenamefont {Esirkepov},
  \citenamefont {Gelfer}, \citenamefont {Bulanov}, \citenamefont {Weber},\ and\
  \citenamefont {Korn}}]{Jirka_PRE2016}%
  \BibitemOpen
  \bibfield  {author} {\bibinfo {author} {\bibfnamefont {M.}~\bibnamefont
  {Jirka}}, \bibinfo {author} {\bibfnamefont {O.}~\bibnamefont {Klimo}},
  \bibinfo {author} {\bibfnamefont {S.~V.}\ \bibnamefont {Bulanov}}, \bibinfo
  {author} {\bibfnamefont {T.~Zh.}\ \bibnamefont {Esirkepov}}, \bibinfo {author}
  {\bibfnamefont {E.}~\bibnamefont {Gelfer}}, \bibinfo {author} {\bibfnamefont
  {S.~S.}\ \bibnamefont {Bulanov}}, \bibinfo {author} {\bibfnamefont
  {S.}~\bibnamefont {Weber}}, \ and\ \bibinfo {author} {\bibfnamefont
  {G.}~\bibnamefont {Korn}},\ }\href {\doibase 10.1103/PhysRevE.93.023207}
  {\bibfield  {journal} {\bibinfo  {journal} {Phys. Rev. E}\ }\textbf {\bibinfo
  {volume} {93}},\ \bibinfo {pages} {023207} (\bibinfo {year}
  {2016})}\BibitemShut {NoStop}%
\bibitem [{\citenamefont {Kostyukov}\ and\ \citenamefont
  {Nerush}(2016)}]{Kostyukov_POP2016}%
  \BibitemOpen
  \bibfield  {author} {\bibinfo {author} {\bibfnamefont {I.~Y.}\ \bibnamefont
  {Kostyukov}}\ and\ \bibinfo {author} {\bibfnamefont {E.~N.}\ \bibnamefont
  {Nerush}},\ }\href {\doibase http://dx.doi.org/10.1063/1.4962567} {\bibfield
  {journal} {\bibinfo  {journal} {Physics of Plasmas}\ }\textbf {\bibinfo
  {volume} {23}},\ \bibinfo {pages} {093119} (\bibinfo {year}
  {2016})}\BibitemShut {NoStop}%
\bibitem [{\citenamefont {Lehmann}\ and\ \citenamefont
  {Spatschek}(2012)}]{Lehmann_PRE}%
  \BibitemOpen
  \bibfield  {author} {\bibinfo {author} {\bibfnamefont {G.}~\bibnamefont
  {Lehmann}}\ and\ \bibinfo {author} {\bibfnamefont {K.~H.}\ \bibnamefont
  {Spatschek}},\ }\href {\doibase 10.1103/PhysRevE.85.056412} {\bibfield
  {journal} {\bibinfo  {journal} {Phys. Rev. E}\ }\textbf {\bibinfo {volume}
  {85}},\ \bibinfo {pages} {056412} (\bibinfo {year} {2012})}\BibitemShut
  {NoStop}%
\bibitem [{\citenamefont {Steiger}\ and\ \citenamefont
  {Woods}(1972)}]{Steiger_PRA1972}%
  \BibitemOpen
  \bibfield  {author} {\bibinfo {author} {\bibfnamefont {A.~D.}\ \bibnamefont
  {Steiger}}\ and\ \bibinfo {author} {\bibfnamefont {C.~H.}\ \bibnamefont
  {Woods}},\ }\href {\doibase 10.1103/PhysRevA.5.1467} {\bibfield  {journal}
  {\bibinfo  {journal} {Phys. Rev. A}\ }\textbf {\bibinfo {volume} {5}},\
  \bibinfo {pages} {1467} (\bibinfo {year} {1972})}\BibitemShut {NoStop}%
\bibitem [{\citenamefont {Zeldovich}(1975)}]{Zeldovich_UFN1975}%
  \BibitemOpen
  \bibfield  {author} {\bibinfo {author} {\bibfnamefont {Y.~B.}\ \bibnamefont
  {Zeldovich}},\ }\href {\doibase 10.1070/PU1975v018n02ABEH001947} {\bibfield
  {journal} {\bibinfo  {journal} {Phys. Usp.}\ }\textbf {\bibinfo {volume}
  {18}},\ \bibinfo {pages} {79} (\bibinfo {year} {1975})}\BibitemShut {NoStop}%
\bibitem [{\citenamefont {Rohrlich}(1965)}]{Rohrlich}%
  \BibitemOpen
  \bibfield  {author} {\bibinfo {author} {\bibfnamefont {R.}~\bibnamefont
  {Rohrlich}},\ }\href@noop {} {\emph {\bibinfo {title} {Classical Charged
  Particles}}}\ (\bibinfo  {publisher} {Addison-Wesley, Reading, MA},\ \bibinfo
  {year} {1965})\BibitemShut {NoStop}%
\bibitem [{\citenamefont {Bulanov}\ \emph {et~al.}(2004)\citenamefont
  {Bulanov}, \citenamefont {Esirkepov}, \citenamefont {Koga},\ and\
  \citenamefont {Tajima}}]{Bulanov_PPR2004}%
  \BibitemOpen
  \bibfield  {author} {\bibinfo {author} {\bibfnamefont {S.~V.}\ \bibnamefont
  {Bulanov}}, \bibinfo {author} {\bibfnamefont {T.~Zh.}\ \bibnamefont
  {Esirkepov}}, \bibinfo {author} {\bibfnamefont {J.}~\bibnamefont {Koga}}, \
  and\ \bibinfo {author} {\bibfnamefont {T.}~\bibnamefont {Tajima}},\ }\href
  {\doibase 10.1134/1.1687021} {\bibfield  {journal} {\bibinfo  {journal}
  {Plasma Physics Reports}\ }\textbf {\bibinfo {volume} {30}},\ \bibinfo
  {pages} {196} (\bibinfo {year} {2004})}\BibitemShut {NoStop}%
\bibitem [{\citenamefont {Bashinov}\ and\ \citenamefont
  {Kim}(2013)}]{Bashinov_PP2013}%
  \BibitemOpen
  \bibfield  {author} {\bibinfo {author} {\bibfnamefont {A.~V.}\ \bibnamefont
  {Bashinov}}\ and\ \bibinfo {author} {\bibfnamefont {A.~V.}\ \bibnamefont
  {Kim}},\ }\href {\doibase 10.1063/1.4835215} {\bibfield  {journal} {\bibinfo
  {journal} {Phys. Plasmas}\ }\textbf {\bibinfo {volume} {20}},\ \bibinfo
  {pages} {113111} (\bibinfo {year} {2013})}\BibitemShut {NoStop}%
\bibitem [{\citenamefont {Zhang}\ \emph {et~al.}(2015)\citenamefont {Zhang},
  \citenamefont {Ridgers},\ and\ \citenamefont {Thomas}}]{Zhang_NJP2015}%
  \BibitemOpen
  \bibfield  {author} {\bibinfo {author} {\bibfnamefont {P.}~\bibnamefont
  {Zhang}}, \bibinfo {author} {\bibfnamefont {C.~P.}\ \bibnamefont {Ridgers}},
  \ and\ \bibinfo {author} {\bibfnamefont {A.~G.~R.}\ \bibnamefont {Thomas}},\
  }\href {http://stacks.iop.org/1367-2630/17/i=4/a=043051} {\bibfield
  {journal} {\bibinfo  {journal} {New Journal of Physics}\ }\textbf {\bibinfo
  {volume} {17}},\ \bibinfo {pages} {043051} (\bibinfo {year}
  {2015})}\BibitemShut {NoStop}%
\bibitem [{\citenamefont {Elkina}\ \emph {et~al.}(2011)\citenamefont {Elkina},
  \citenamefont {Fedotov}, \citenamefont {Kostyukov}, \citenamefont {Legkov},
  \citenamefont {Narozhny}, \citenamefont {Nerush},\ and\ \citenamefont
  {Ruhl}}]{Elkina_PRSTAB2011}%
  \BibitemOpen
  \bibfield  {author} {\bibinfo {author} {\bibfnamefont {N.~V.}\ \bibnamefont
  {Elkina}}, \bibinfo {author} {\bibfnamefont {A.~M.}\ \bibnamefont {Fedotov}},
  \bibinfo {author} {\bibfnamefont {I.~Y.}\ \bibnamefont {Kostyukov}}, \bibinfo
  {author} {\bibfnamefont {M.~V.}\ \bibnamefont {Legkov}}, \bibinfo {author}
  {\bibfnamefont {N.~B.}\ \bibnamefont {Narozhny}}, \bibinfo {author}
  {\bibfnamefont {E.~N.}\ \bibnamefont {Nerush}}, \ and\ \bibinfo {author}
  {\bibfnamefont {H.}~\bibnamefont {Ruhl}},\ }\href {\doibase
  10.1103/PhysRevSTAB.14.054401} {\bibfield  {journal} {\bibinfo  {journal}
  {Phys. Rev. ST Accel. Beams}\ }\textbf {\bibinfo {volume} {14}},\ \bibinfo
  {pages} {054401} (\bibinfo {year} {2011})}\BibitemShut {NoStop}%
\bibitem [{\citenamefont {Nerush}\ \emph
  {et~al.}(2011{\natexlab{b}})\citenamefont {Nerush}, \citenamefont
  {Bashmakov},\ and\ \citenamefont {Kostyukov}}]{Nerush_POP2011}%
  \BibitemOpen
  \bibfield  {author} {\bibinfo {author} {\bibfnamefont {E.~N.}\ \bibnamefont
  {Nerush}}, \bibinfo {author} {\bibfnamefont {V.~F.}\ \bibnamefont
  {Bashmakov}}, \ and\ \bibinfo {author} {\bibfnamefont {I.~Y.}\ \bibnamefont
  {Kostyukov}},\ }\href {\doibase 10.1063/1.3624481} {\bibfield  {journal}
  {\bibinfo  {journal} {Phys. Plasmas}\ }\textbf {\bibinfo {volume} {18}},\
  \bibinfo {pages} {083107} (\bibinfo {year} {2011}{\natexlab{b}})}\BibitemShut
  {NoStop}%
\bibitem [{\citenamefont {Fedotov}\ \emph {et~al.}(2010)\citenamefont
  {Fedotov}, \citenamefont {Narozhny}, \citenamefont {Mourou},\ and\
  \citenamefont {Korn}}]{Fedotov_PRL2010}%
  \BibitemOpen
  \bibfield  {author} {\bibinfo {author} {\bibfnamefont {A.~M.}\ \bibnamefont
  {Fedotov}}, \bibinfo {author} {\bibfnamefont {N.~B.}\ \bibnamefont
  {Narozhny}}, \bibinfo {author} {\bibfnamefont {G.}~\bibnamefont {Mourou}}, \
  and\ \bibinfo {author} {\bibfnamefont {G.}~\bibnamefont {Korn}},\ }\href
  {\doibase 10.1103/PhysRevLett.105.080402} {\bibfield  {journal} {\bibinfo
  {journal} {Phys. Rev. Lett.}\ }\textbf {\bibinfo {volume} {105}},\ \bibinfo
  {pages} {080402} (\bibinfo {year} {2010})}\BibitemShut {NoStop}%
\bibitem [{\citenamefont {Fedotov}\ \emph {et~al.}(2014)\citenamefont
  {Fedotov}, \citenamefont {Elkina}, \citenamefont {Gelfer}, \citenamefont
  {Narozhny},\ and\ \citenamefont {Ruhl}}]{Fedotov_PRA2014}%
  \BibitemOpen
  \bibfield  {author} {\bibinfo {author} {\bibfnamefont {A.~M.}\ \bibnamefont
  {Fedotov}}, \bibinfo {author} {\bibfnamefont {N.~V.}\ \bibnamefont {Elkina}},
  \bibinfo {author} {\bibfnamefont {E.~G.}\ \bibnamefont {Gelfer}}, \bibinfo
  {author} {\bibfnamefont {N.~B.}\ \bibnamefont {Narozhny}}, \ and\ \bibinfo
  {author} {\bibfnamefont {H.}~\bibnamefont {Ruhl}},\ }\href {\doibase
  10.1103/PhysRevA.90.053847} {\bibfield  {journal} {\bibinfo  {journal} {Phys.
  Rev. A}\ }\textbf {\bibinfo {volume} {90}},\ \bibinfo {pages} {053847}
  (\bibinfo {year} {2014})}\BibitemShut {NoStop}%
\bibitem [{\citenamefont {Kirk}\ \emph {et~al.}(2009)\citenamefont {Kirk},
  \citenamefont {Bell},\ and\ \citenamefont {Arka}}]{Kirk_PPCF2009}%
  \BibitemOpen
  \bibfield  {author} {\bibinfo {author} {\bibfnamefont {J.~G.}\ \bibnamefont
  {Kirk}}, \bibinfo {author} {\bibfnamefont {A.~R.}\ \bibnamefont {Bell}}, \
  and\ \bibinfo {author} {\bibfnamefont {I.}~\bibnamefont {Arka}},\ }\href
  {http://stacks.iop.org/0741-3335/51/i=8/a=085008} {\bibfield  {journal}
  {\bibinfo  {journal} {Plasma Physics and Controlled Fusion}\ }\textbf
  {\bibinfo {volume} {51}},\ \bibinfo {pages} {085008} (\bibinfo {year}
  {2009})}\BibitemShut {NoStop}%
\bibitem [{\citenamefont {Tamburini}\ \emph {et~al.}(2010)\citenamefont
  {Tamburini}, \citenamefont {Pegoraro}, \citenamefont {Piazza}, \citenamefont
  {Keitel},\ and\ \citenamefont {Macchi}}]{Tamburini_NJP}%
  \BibitemOpen
  \bibfield  {author} {\bibinfo {author} {\bibfnamefont {M.}~\bibnamefont
  {Tamburini}}, \bibinfo {author} {\bibfnamefont {F.}~\bibnamefont {Pegoraro}},
  \bibinfo {author} {\bibfnamefont {A.~D.}\ \bibnamefont {Piazza}}, \bibinfo
  {author} {\bibfnamefont {C.~H.}\ \bibnamefont {Keitel}}, \ and\ \bibinfo
  {author} {\bibfnamefont {A.}~\bibnamefont {Macchi}},\ }\href {\doibase
  10.1088/1367-2630/12/12/123005} {\bibfield  {journal} {\bibinfo  {journal}
  {New Journal of Physics}\ }\textbf {\bibinfo {volume} {12}},\ \bibinfo
  {pages} {123005} (\bibinfo {year} {2010})}\BibitemShut {NoStop}%
\bibitem [{\citenamefont {Bulanov}\ \emph {et~al.}(2011)\citenamefont
  {Bulanov}, \citenamefont {Esirkepov}, \citenamefont {Kando}, \citenamefont
  {Koga},\ and\ \citenamefont {Bulanov}}]{Bulanov_PRE2011}%
  \BibitemOpen
  \bibfield  {author} {\bibinfo {author} {\bibfnamefont {S.~V.}\ \bibnamefont
  {Bulanov}}, \bibinfo {author} {\bibfnamefont {T.~Zh.}\ \bibnamefont
  {Esirkepov}}, \bibinfo {author} {\bibfnamefont {M.}~\bibnamefont {Kando}},
  \bibinfo {author} {\bibfnamefont {J.~K.}\ \bibnamefont {Koga}}, \ and\
  \bibinfo {author} {\bibfnamefont {S.~S.}\ \bibnamefont {Bulanov}},\ }\href
  {\doibase 10.1103/PhysRevE.84.056605} {\bibfield  {journal} {\bibinfo
  {journal} {Phys. Rev. E}\ }\textbf {\bibinfo {volume} {84}},\ \bibinfo
  {pages} {056605} (\bibinfo {year} {2011})}\BibitemShut {NoStop}%
\bibitem [{\citenamefont {Bayer}\ \emph {et~al.}(1973)\citenamefont {Bayer},
  \citenamefont {Katkov},\ and\ \citenamefont {Fadin}}]{BayerKatkov}%
  \BibitemOpen
  \bibfield  {author} {\bibinfo {author} {\bibfnamefont {V.~N.}\ \bibnamefont
  {Bayer}}, \bibinfo {author} {\bibfnamefont {V.~M.}\ \bibnamefont {Katkov}}, \
  and\ \bibinfo {author} {\bibfnamefont {V.~S.}\ \bibnamefont {Fadin}},\
  }\href@noop {} {\emph {\bibinfo {title} {Radiation of the Relativistic
  Electrons}}}\ (\bibinfo  {publisher} {Atomizdat, Moscow},\ \bibinfo {year}
  {1973})\BibitemShut {NoStop}%
\bibitem [{\citenamefont {Nikishov}(1985)}]{Nikishov_85}%
  \BibitemOpen
  \bibfield  {author} {\bibinfo {author} {\bibfnamefont {A.}~\bibnamefont
  {Nikishov}},\ }\href {\doibase 10.1007/BF01120143} {\bibfield  {journal}
  {\bibinfo  {journal} {Journal of Soviet Laser Research}\ }\textbf {\bibinfo
  {volume} {6}},\ \bibinfo {pages} {619} (\bibinfo {year} {1985})}\BibitemShut
  {NoStop}%
\bibitem [{\citenamefont {Ritus}(1985)}]{Ritus_85}%
  \BibitemOpen
  \bibfield  {author} {\bibinfo {author} {\bibfnamefont {V.}~\bibnamefont
  {Ritus}},\ }\href {\doibase 10.1007/BF01120220} {\bibfield  {journal}
  {\bibinfo  {journal} {Journal of Soviet Laser Research}\ }\textbf {\bibinfo
  {volume} {6}},\ \bibinfo {pages} {497} (\bibinfo {year} {1985})}\BibitemShut
  {NoStop}%
\bibitem [{\citenamefont {Gonoskov}\ \emph {et~al.}(2015)\citenamefont
  {Gonoskov}, \citenamefont {Bastrakov}, \citenamefont {Efimenko},
  \citenamefont {Ilderton}, \citenamefont {Marklund}, \citenamefont {Meyerov},
  \citenamefont {Muraviev}, \citenamefont {Sergeev}, \citenamefont {Surmin},\
  and\ \citenamefont {Wallin}}]{Gonoskov_PRE2015}%
  \BibitemOpen
  \bibfield  {author} {\bibinfo {author} {\bibfnamefont {A.}~\bibnamefont
  {Gonoskov}}, \bibinfo {author} {\bibfnamefont {S.}~\bibnamefont {Bastrakov}},
  \bibinfo {author} {\bibfnamefont {E.}~\bibnamefont {Efimenko}}, \bibinfo
  {author} {\bibfnamefont {A.}~\bibnamefont {Ilderton}}, \bibinfo {author}
  {\bibfnamefont {M.}~\bibnamefont {Marklund}}, \bibinfo {author}
  {\bibfnamefont {I.}~\bibnamefont {Meyerov}}, \bibinfo {author} {\bibfnamefont
  {A.}~\bibnamefont {Muraviev}}, \bibinfo {author} {\bibfnamefont
  {A.}~\bibnamefont {Sergeev}}, \bibinfo {author} {\bibfnamefont
  {I.}~\bibnamefont {Surmin}}, \ and\ \bibinfo {author} {\bibfnamefont
  {E.}~\bibnamefont {Wallin}},\ }\href {\doibase 10.1103/PhysRevE.92.023305}
  {\bibfield  {journal} {\bibinfo  {journal} {Phys. Rev. E}\ }\textbf {\bibinfo
  {volume} {92}},\ \bibinfo {pages} {023305} (\bibinfo {year}
  {2015})}\BibitemShut {NoStop}%
\bibitem [{\citenamefont {Landau}\ and\ \citenamefont
  {Lifshitz}(1975)}]{Landau2}%
  \BibitemOpen
  \bibfield  {author} {\bibinfo {author} {\bibfnamefont {L.~D.}\ \bibnamefont
  {Landau}}\ and\ \bibinfo {author} {\bibfnamefont {E.~M.}\ \bibnamefont
  {Lifshitz}},\ }\href@noop {} {\emph {\bibinfo {title} {The classical theory
  of fields}}}\ (\bibinfo  {publisher} {Elsevier, Oxford},\ \bibinfo {year}
  {1975})\BibitemShut {NoStop}%
\bibitem [{\citenamefont {Bastrakov}\ \emph {et~al.}(2012)\citenamefont
  {Bastrakov}, \citenamefont {Donchenko}, \citenamefont {Gonoskov},
  \citenamefont {Efimenko}, \citenamefont {Malyshev}, \citenamefont {Meyerov},\
  and\ \citenamefont {Surmin}}]{Bastrakov_JCS2012}%
  \BibitemOpen
  \bibfield  {author} {\bibinfo {author} {\bibfnamefont {S.}~\bibnamefont
  {Bastrakov}}, \bibinfo {author} {\bibfnamefont {R.}~\bibnamefont
  {Donchenko}}, \bibinfo {author} {\bibfnamefont {A.}~\bibnamefont {Gonoskov}},
  \bibinfo {author} {\bibfnamefont {E.}~\bibnamefont {Efimenko}}, \bibinfo
  {author} {\bibfnamefont {A.}~\bibnamefont {Malyshev}}, \bibinfo {author}
  {\bibfnamefont {I.}~\bibnamefont {Meyerov}}, \ and\ \bibinfo {author}
  {\bibfnamefont {I.}~\bibnamefont {Surmin}},\ }\href {\doibase
  10.1016/j.jocs.2012.08.012} {\bibfield  {journal} {\bibinfo  {journal} {J.
  Comput. Sci.}\ }\textbf {\bibinfo {volume} {3}},\ \bibinfo {pages} {474 }
  (\bibinfo {year} {2012})}\BibitemShut {NoStop}%
\end{thebibliography}%
\end{document}